\documentclass[%
 reprint,
superscriptaddress,
showpacs,preprintnumbers,
nofootinbib,
 amsmath,amssymb, 
 aps,
 prd,
 longbibliography,
]{revtex4-1}

\usepackage{cancel}
\usepackage{accents}
\usepackage{mciteplus,slashed}
\usepackage{amssymb,cancel,amsmath}
\usepackage{dcolumn}
\usepackage{bm}
\usepackage[caption=false]{subfig}
\usepackage{appendix}
\usepackage{physics}
\usepackage{booktabs}
\usepackage{feynmp-auto}
\usepackage[T1]{fontenc}	
\usepackage{csvsimple}
\usepackage{hyperref}
\usepackage[section]{placeins}
\usepackage[capitalise]{cleveref}
\usepackage{booktabs}
\usepackage{graphicx}
\usepackage{mathrsfs}
\usepackage[caption=false]{subfig}
\graphicspath{{Figures/}}

\AtBeginDocument{
\heavyrulewidth=.08em
\lightrulewidth=.05em
\cmidrulewidth=.03em
\belowrulesep=.65ex
\belowbottomsep=0pt
\aboverulesep=.4ex
\abovetopsep=0pt
\cmidrulesep=\doublerulesep
\cmidrulekern=.5em
\defaultaddspace=.5em
}
\usepackage[dvipsnames]{xcolor}
\usepackage[normalem]{ulem}
\usepackage{fontawesome} 

\def\e{\mathrm{e}}
\def\Ldet{\ell}
\def\Adet{A_\perp}

\def\Pvis{P_{\rm vis}}
\def\deg{^{\circ}}

\graphicspath{{Figures/}}

\begin{document}

\title{Neutrino Portals, Terrestrial Upscattering, and Atmospheric Neutrinos }


\author{R. Andrew Gustafson}
\email{gustafr@vt.edu}
\affiliation{Center for Neutrino Physics, Department of Physics,
Virginia Tech University, Blacksburg, VA 24061, USA}

\author{Ryan Plestid}
\email{rpl225@uky.edu}
\affiliation{Department of Physics and Astronomy, University of Kentucky  Lexington, KY 40506, USA}
\affiliation{Theoretical Physics Department, Fermilab, Batavia, IL 60510,USA}
\preprint{FERMILAB-PUB-22-250-V}

\author{Ian M. Shoemaker}
\email{shoemaker@vt.edu}
\affiliation{Center for Neutrino Physics, Department of Physics,
Virginia Tech University, Blacksburg, VA 24061, USA}
 
\begin{abstract}
     {\centering{\href{https://github.com/ryanplestid/Atmo-UpScatt}{\large\color{BlueViolet}\faGithub}}  \\}  
     We consider the upscattering of atmospheric neutrinos in the interior of the Earth producing heavy neutral leptons (HNLs) which subsequently decay inside large volume  detectors (e.g. Super-Kamiokande or DUNE). We compute the flux of upscattered HNLs arriving at a detector, and the resultant event rate of visible decay products. Using Super-Kamiokande's atmospheric neutrino dataset we find new leading constraints for dipole couplings to any flavor with HNL masses between roughly 10 MeV and 100 MeV. For mass mixing with tau neutrinos, we probe new parameter space near HNL masses of $\sim 20$ MeV with prospects for substantial future improvements. We also discuss prospects at future experiments such as DUNE, JUNO, and Hyper-Kamiokande. 
\end{abstract}

 \maketitle 
 
 \section{Introduction \label{Introduction}}
 Heavy neutral leptons (HNLs) are well motivated extensions of the Standard Model (SM). They appear ubiquitously in dark sector models, and are especially important because their coupling to neutrinos represents one of three unique renormalizable ``portals'' between a generic dark sector and the Standard Model \cite{Patt:2006fw,Batell:2009di,Beacham:2019nyx,Agrawal:2021dbo}. They are natural partners to left-handed neutrinos, reflecting a matching of chiral degrees of freedom observed among all other SM fermions, and their non-observation is easily explained due to their being a SM gauge singlet. They are further motivated by anomaly-cancellation arguments that become essential in models with new gauge groups  e.g.\ a gauged $B-L$ \cite{Alvarez-Gaume:1983ihn,Montero:2007cd}. HNLs appear as necessary ingredients in certain grand unified theories (GUTs) e.g.\ SO(10) \cite{Fritzsch:1974nn}, and they are intimately connected to neutrino masses.\footnote{The Zee-Babu mechanism \cite{Babu:1988ki,Zee:1985id,Herrero-Garcia:2014hfa} is a notable counter example in which neutrino masses are induced via new scalars rather than fermions.} 
 
 Despite their strong theoretical motivation, there is no model-independent prediction for the HNL mass scale. HNLs may be $O(\rm eV)$ in mass (i.e.\ sterile neutrinos) and connected to neutrino masses via a naive Dirac mass mechanism (like all other SM fermions), in which case they are most easily searched for using short-baseline oscillation experiments or cosmological observables. Alternatively, neutrino masses may be generated by a type-I \cite{Minkowski:1977sc,Yanagida:1979as,Gell-Mann:1979vob} seesaw mechanism each of which leads to different expected mass scales for different Yukawa couplings.
 
 The ubiquity of HNLs in generic models of a dark sector, and their unconstrained mass range therefore motivates a broad search strategy that targets many decades of HNL mass parameter space, ranging from the eV to the GeV (or even TeV) scale \cite{Arguelles:2019ziu,Liventsev:2013zz,Asaka:2005pn,Asaka:2005an,Atre:2009rg,Johnson:1997cj,Levy:2018dns,Formaggio:1998zn,Gorbunov:2007ak,Drewes:2013gca,Bondarenko:2018ptm,Boyarsky:2009ix,Berryman:2019dme,Ballett:2019bgd}. In this work we will demonstrate that large volume detectors can efficiently search for HNLs via their decays by leveraging atmospheric neutrino upscattering inside the Earth. This search strategy is ideally suited to HNL decay lengths, $\lambda$, satisfying $ 10~{\rm m}\lesssim \lambda \lesssim 6000 ~{\rm km}$ with the upper limit set by the Earth's radius. This complements fixed target ``beam dump'' experiments and missing energy searches which typically lose sensitivity as the HNL decay length becomes much longer than the experimental apparatus, which tend to range from  10s to 100s of meters (much shorter than the 1000s of km that characterize the Earth's radius). We derive new and leading constraints for HNL masses between $\sim 10~ \rm MeV$ and $200 ~\rm MeV$, a mass range which has interesting implications for models that address the Hubble tension \cite{Gelmini:2019deq}.
 
 Much of the literature on HNLs focuses on the aforementioned renormalizable coupling between HNLs and active neutrinos available in the SM. This is the so-called mass-mixing portal, which results in a small admixture of HNL contamination among the active neutrinos
 \begin{equation}
    \nu_\alpha = U_{\alpha N} N + \sum_{i=1}^3 U_{\alpha i}\nu_i~, 
 \end{equation}
where $U$ is the mixing angle, and where $\alpha \in \{e, \mu, \tau\}$ labels the active neutrino species in the flavor basis. This then induces a transition matrix elements within the weak current, e.g.\  $\mathcal{L}_{\rm int} \supset U_{\alpha N} \bar{N}_i \gamma_\mu P_L \nu_\alpha J^\mu$. 
 
Above the weak scale the mass-mixing portal is relevant in the Wilsonian sense, however below the weak scale the mixing angle accompanies an irrelevant dimension-6 operator in the 4-Fermi effective theory that governs low-energy neutrino phenomenology. Despite being Wilsonian-irrelevant, the mass-mixing portal can still be efficiently probed at low energies because of its strength \emph{relative} to SM neutrino interactions which proceed through the same dimension-6 contact operators.

 There is, however, one unique portal that is dimension-5 and so can come to dominate over SM weak currents at low energies even if it is sub-dominant at high energies. This is the so-called ``dipole portal'' which first received substantial attention in the context of the MiniBooNE and LSND anomalies \cite{Gninenko:2009ks,Gninenko:2010pr,McKeen:2010rx,Masip:2011qb,Masip:2012ke}. The authors of \cite{Coloma:2017ppo} further pointed out interesting ``double-bang'' phenomenology that could be probed in experiments such as IceCube. Ref.\ \cite{Magill:2018jla} initiated a broad study of the relevant parameter space for a dipole portal, and this was recently complemented by a thorough analysis of low-energy and cosmological phenomena \cite{Brdar:2020quo}. Since these early studies, the viable parameter space for a neutrino dipole portal has received considerable attention \cite{Shoemaker:2018vii,Arguelles:2018mtc,Hostert:2019iia,Fischer:2019fbw,Coloma:2019htx,Schwetz:2020xra,Arina:2020mxo,Shoemaker:2020kji,Abdullahi:2020nyr,Shakeri:2020wvk,Atkinson:2021rnp,Cho:2021yxk,Kim:2021lun,Arguelles:2021dqn,Ismail:2021dyp,Miranda:2021kre,Bolton:2021pey,Jodlowski:2020vhr}, and has persisted as a potential explanation of the MiniBooNE excess \cite{Fischer:2019fbw,Vergani:2021tgc}. 
 
 The interaction Lagrangian for the dipole portal is conventionally taken to be
\begin{equation}    
    \label{dip-lag}
    \mathcal{L}_{\rm int} \supset \sum_\alpha d_\alpha F^{\mu\nu}\bar{N} \sigma_{\mu\nu} P_L \nu_\alpha~. 
\end{equation}
 where $d_\alpha$ is the (flavor dependent) transition dipole moment between $\nu_\alpha$ and the singlet fermion $N$. 
 In complete generality one could consider a linear combination of magnetic and electric transition dipole portals (see \cite{Sierra:2021say} for a recent discussion and \cite{Balantekin:2018ukw,deGouvea:2021rpa} for related work in the context of angular distributions in HNL decays). It suffices, however, to consider only the magnetic dipole portal as a simplified model in the majority of parameter space, and we restrict our attention to this case here. 
 
 Constructing UV-completions that yield sizeable dipole operators is a non-trivial model building task. One constraint stems from neutrino masses, since loop diagrams involving a photon insertion on the incoming and outgoing neutrino can alter neutrino textures. This can be avoided if $N$ is a Dirac or psuedo-Dirac fermion \cite{Magill:2018jla}. Ref.\ \cite{Brdar:2020quo} discusses possible UV completions connected to leptoquarks and recent B-anomalies, and other models have been discussed in \cite{Babu:2021jnu}. In this paper we work purely at the level of low-energy Lagrangian \cref{dip-lag} and remain agnostic to the UV-origin of the dipole portal. For the results show below, we consider N to be a Dirac fermion.  From a phenomenological standpoint, the primary difference between Dirac and Majorana HNLs is that the Majorana decay length is half of the Dirac decay length \cite{Plestid:2020vqf}.  We will show later that the lower bounds we obtain on the model are independent of decay length, so we expect similar bounds for Majorana HNLs.

 Finally, we note that HNLs appear generically in dark sector models and that it is natural to consider models in which there are additional light degrees of freedom. While observational evidence demands that low-mass dark sector particles are weakly coupled to the SM, they needn't be weakly coupled to one another and it is consistent and arguably generic for there to exist complicated dark sector dynamics (e.g.~\cite{Batell:2009di}). A simple example is a model with three HNLs and one massive $Z'$ that interacts with the HNLs via $O(1)$ couplings, but is secluded from the SM except for small kinetic mixing terms with e.g.\ the SM photon (e.g.~\cite{Bertuzzo:2018itn,Ballett:2019pyw,Ballett:2019cqp,Abdullahi:2020nyr}). New dark sector dynamics can modify e.g.\ decay lengths and couplings to nuclei, however, while details can change, the same basic phenomenology proceeds: neutrinos scatter on nuclei and produce HNLs, and those HNLs decay inside detectors producing a visible decay signature. This is illustrative of the fact that \emph{many} neutrino-portal dark sectors may be efficiently probed via terrestrial upscattering provided the dark sector has a long-lived visibly decaying particle in its spectrum. 

\begin{figure}
    \includegraphics[height=1\linewidth]{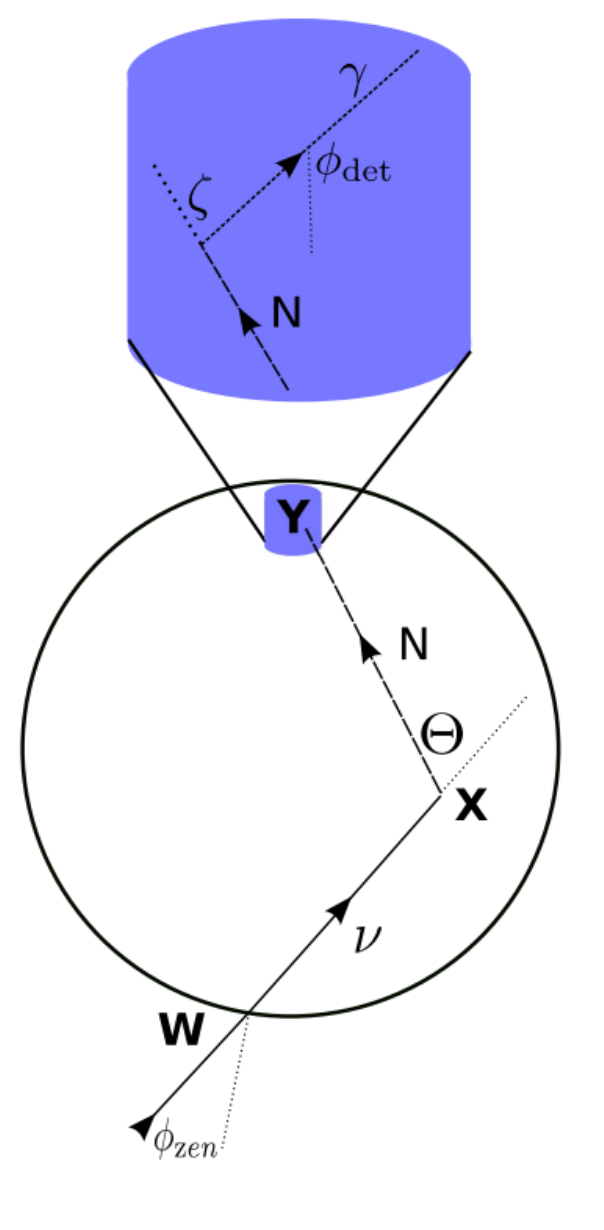}
    \caption{Schematic of upscattering within the Earth and decaying in the detector. A neutrino enters at \textbf{W} with angle $\phi_z$ relative to the vertical. The neutrino scatters into an HNL at \textbf{X} with scattering angle $\Theta$. The HNL reaches the detector at \textbf{Y}, decays within the detector into a neutrino and a visible photon. The photon is emitted with angle $\zeta$ relative to the HNL, and detected with angle $\phi_{det}$ relative to the detector. \label{cartoon}}
\end{figure}

In this work we derive new constraints on neutrino portals using existing data by leveraging atmospheric neutrino upscattering (see \cref{cartoon} for a pictorial description), with our results summarized in \cref{money-plot}. Recent work \cite{Plestid:2020vqf,Plestid:2020ssy} has identified cosmic ray showers as a potentially useful source of HNLs, however our search strategy differs in that the HNLs are produced via volumetric upscattering within the Earth, rather than being produced directly via meson decays in a cosmic rays shower. Because of the large volume of the Earth, this search strategy is ideally suited for regions of parameter space in which the HNL decay length is smaller than, or comparable to, the radius of the Earth. The signature of interest is the through-going decay of an HNL into some visible SM degrees of freedom and the production mechanism is $\nu A \rightarrow N X$ with $A$ some SM particle (typically a nucleus) that is naturally abundant within the Earth's mantle and/or core. The treatment of atmospheric neutrino upscattering is considerably more complicated than solar neutrino upscattering which was pursued previously by one of us in \cite{Plestid:2020vqf,Plestid:2020ssy}. The largest technical challenge is that atmospheric neutrinos oscillate over $O({\rm km})$ length scales. This demands a detailed treatment that includes electron number density profiles along arbitrary line segments within the Earth. In this paper we develop a Monte Carlo routine that is capable of computing the expected event yield inside a detector taking into account all relevant physical details. 

The rest of this paper is organized as follows. In \cref{upscattering} we discuss the upscattering of atmospheric neutrinos. This includes a discussion of neutrino oscillations, atmospheric neutrino intensities, and relevant formulae for upscattering cross sections (both coherent and incoherent). Details of the numerical implementation are deferred to \cref{Monte-Carlo}. In \cref{decays} we discuss the visible signatures of through-going HNLs in large volume detectors. For the dipole portal the signal is always a broad spectrum of photons, whereas for the mass-mixing portal branching ratios vary depending on the HNL mass. Next in \cref{results} we derive new constraints on neutrino-portal couplings to HNLs using Super-Kamiokande (SK) data and discuss potential improvements with both Hyper-Kamiokande (HK) and DUNE in \cref{dedicated}. Finally in \cref{conclusions} we summarize our findings and discuss potential future directions. 



\section{Upscattering in the Earth \label{upscattering} }

  \begin{figure*}
    \subfloat[Constraints assuming $d_\mu=d_\tau=0$ \label{subfig-e}]{%
      \includegraphics[width=0.475\textwidth]{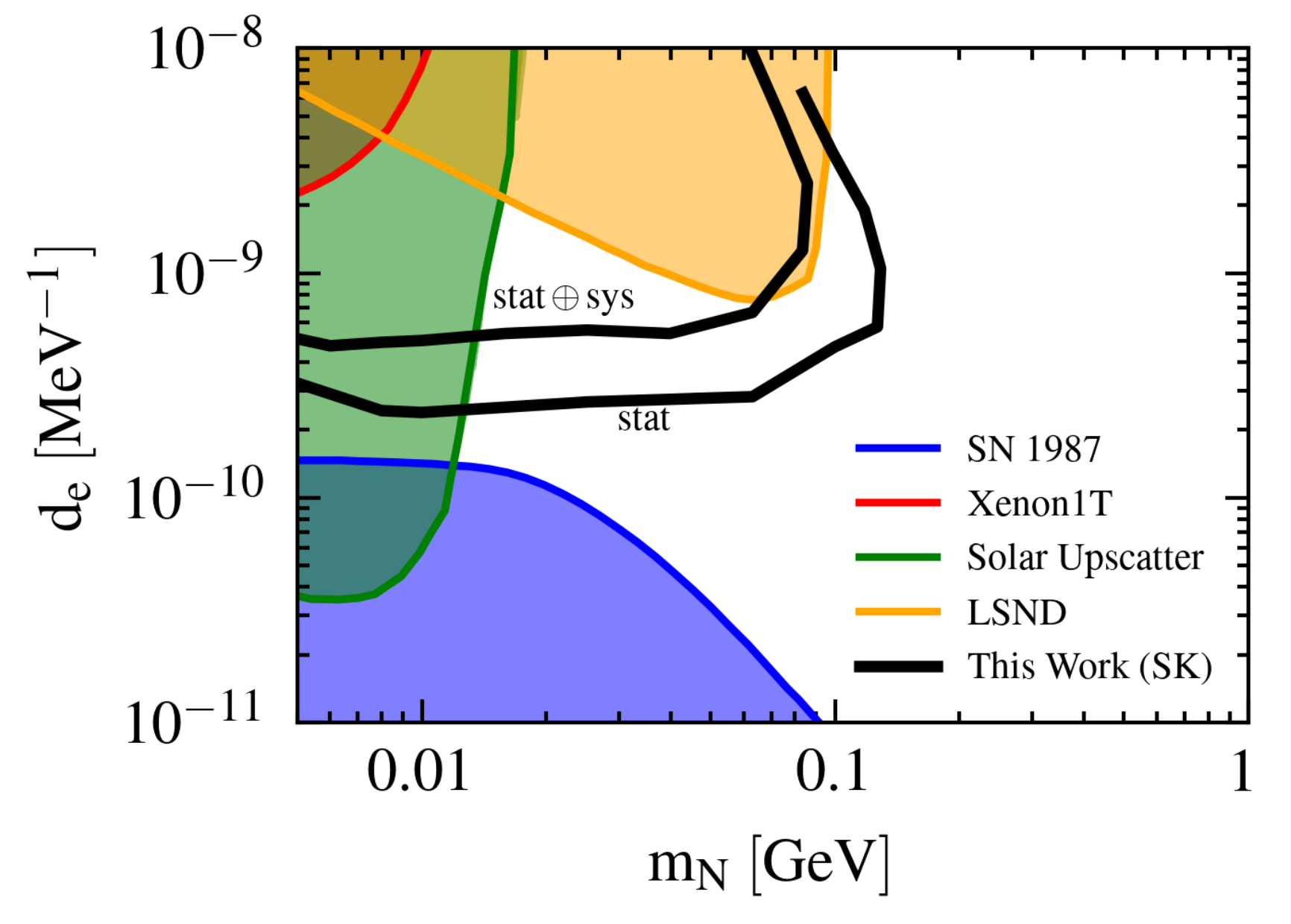}
    }
    \hfill
    \subfloat[Constraints assuming $d_e=d_\tau=0$ \label{subfig-mu}]{%
      \includegraphics[width=0.475\linewidth]{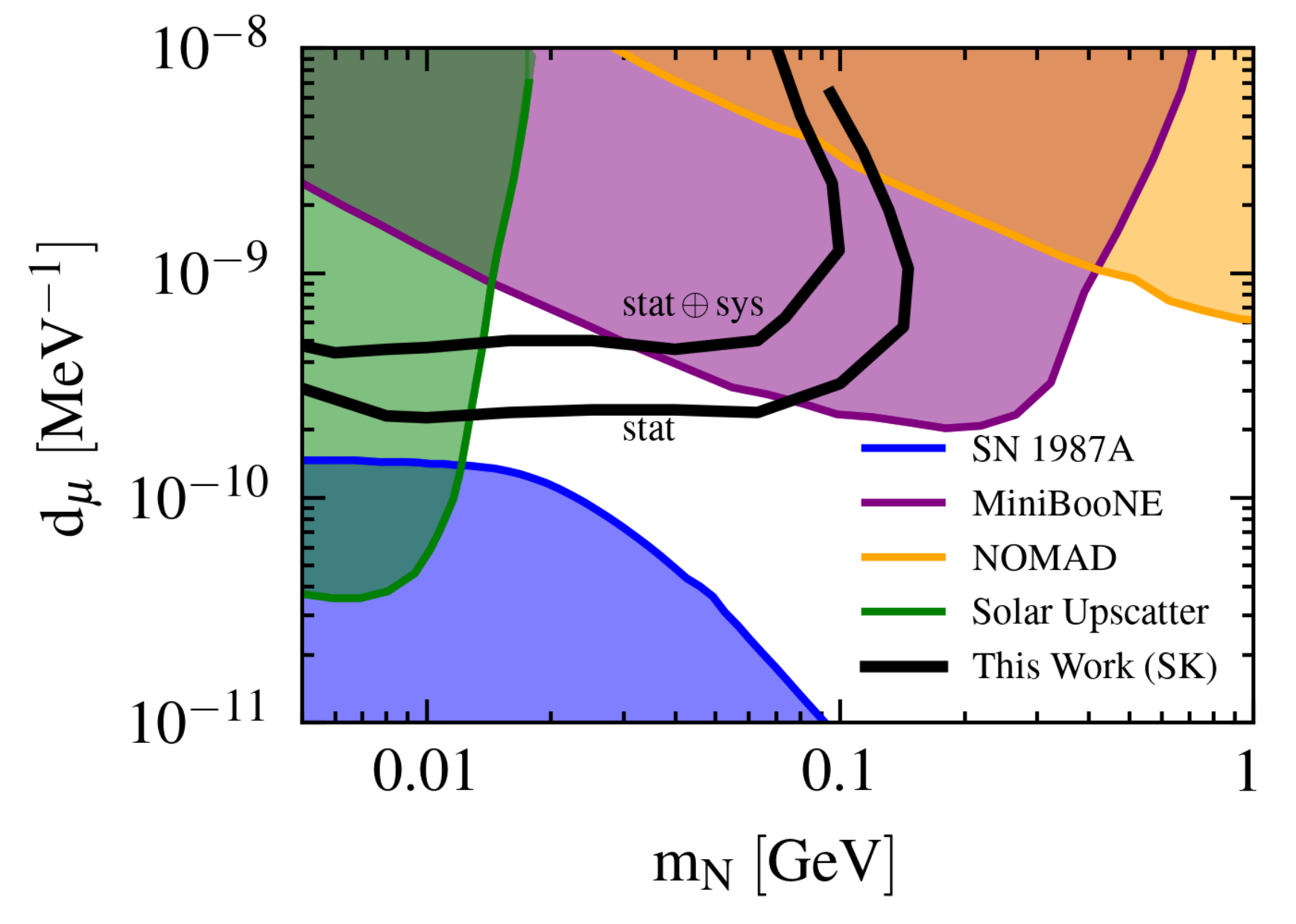}
    }

    \subfloat[Constraints assuming $d_e=d_\tau=0$ \label{subfig-tau}]{%
      \includegraphics[width=0.475\textwidth]{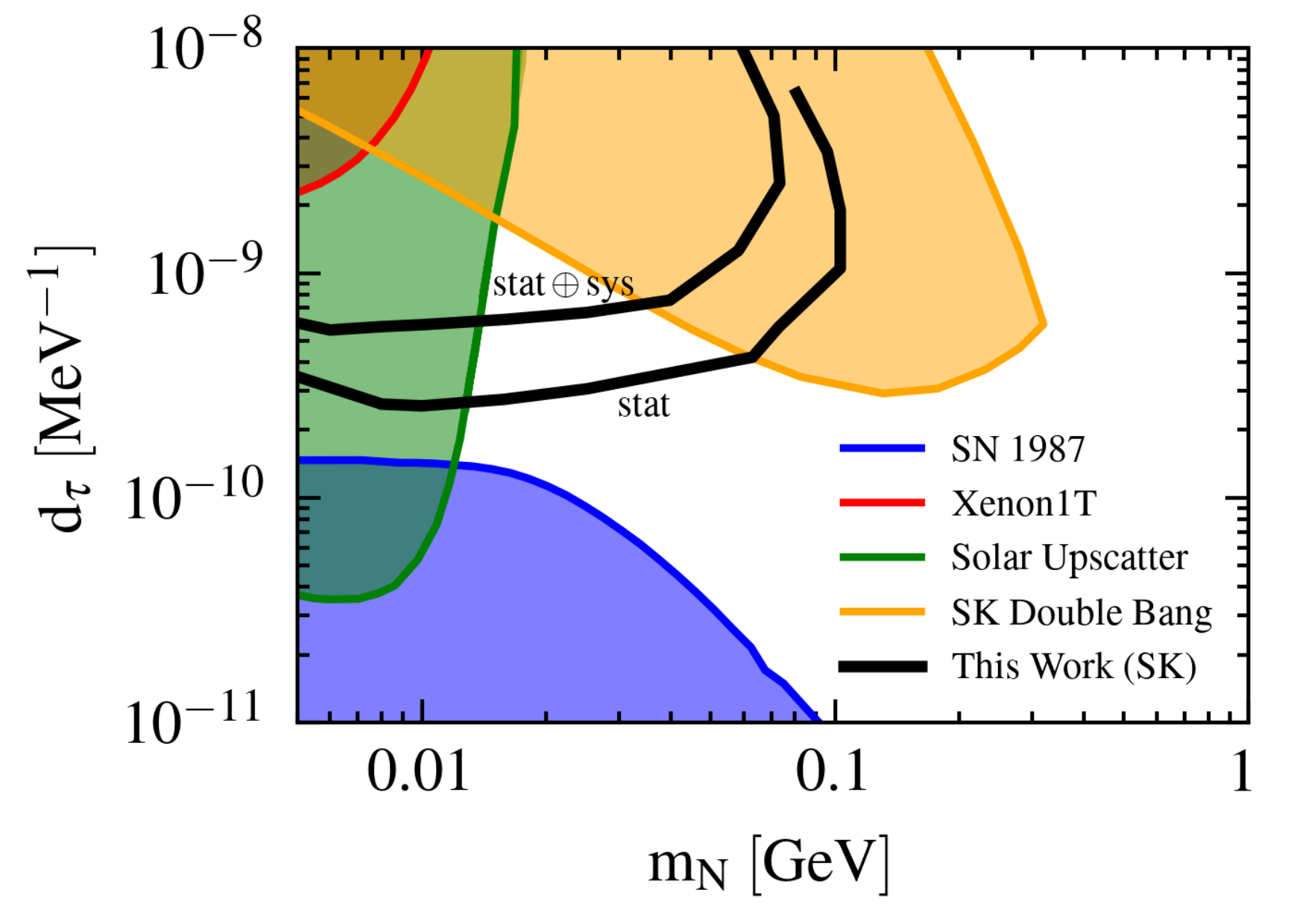}
      }
    \hfill
    \subfloat[Constraints assuming $U_{eN}=U_{\mu N}=0$ \label{subfig-U}]{ \includegraphics[width = 0.475\linewidth]{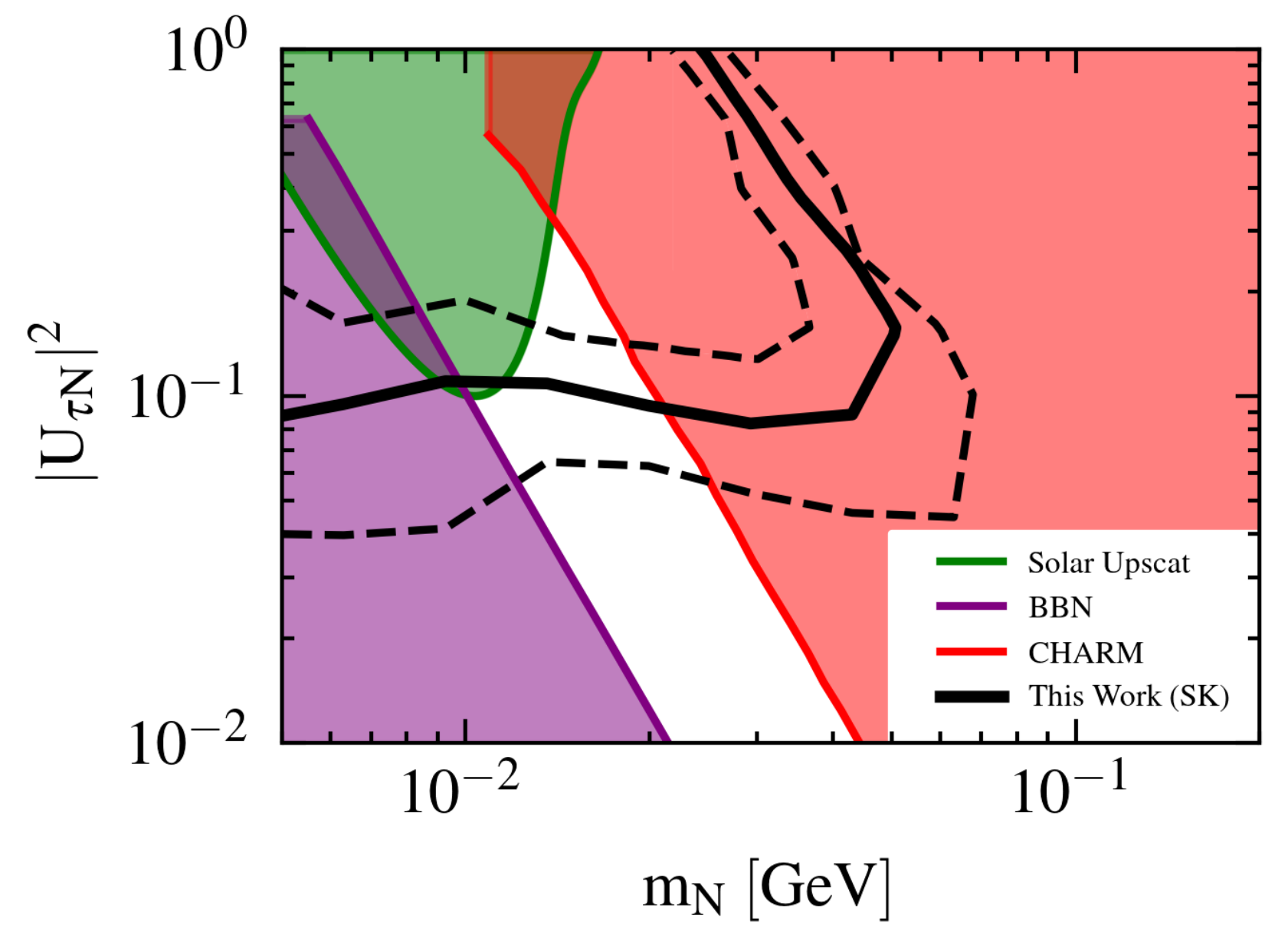}
    }
    
        \caption{Comparison of our dipole-portal limits (a)-(c) and mass-mixing limit for $|U_{\tau N}|^2$ (d) vs. existing limits derived in the literature (for a helpful compilation of \emph{projected} sensitivities see \cite{Schwetz:2020xra}). We have included two bounds from Super-Kamiokande for the dipole portal; one where $\sigma_{\rm{sys}} = 0$ and another where $\sigma_{\rm{sys}}$ is 5\% of the background events. Notice that due to atmospheric neutrino oscillations averaged over the interior of the Earth our constraints are flavor independent up to an $O(1)$ factor. Constraints from NOMAD ($\mu$-only) are taken from~\cite{NOMAD:1997pcg,Coloma:2017ppo}, MiniBooNE ($\mu$-only), supernova bounds (assumed flavor democratic), and LSND constraints are taken from \cite{Magill:2018jla}, and solar constraints are taken from \cite{Plestid:2020vqf}. The $\tau$-only mass-mixing constraints are taken from \cite{Plestid:2020ssy,PhysRevLett.127.121801,Orloff:2002de,Boiarska:2021yho,Atkinson:2021rnp, sabti2020extended}. The dashed lines in (d) are meant to illustrate the theoretical uncertainty in the production rate due to incoherent scattering on nuclei. The lower (upper) dashed line is where we double (halve) the contribution due to incoherent scattering. The coherent contribution (which is nearly free of nuclear uncertainties) guarantees an irreducible flux atmospheric upscattered HNLs. \label{money-plot}}
  \end{figure*}

Atmospheric neutrinos supply a broadband spectrum of electron and muon flavor neutrinos ranging from $\sim 100$ MeV up to $\sim$100s of GeV. At these energies the neutrinos pass through the Earth without scattering, but do undergo substantial flavor oscillations that depend non-trivially on the matter profile encountered by the neutrinos in transit. At a typical point inside the Earth, this results in a quasi-isotropic intensity of neutrinos with a $O(1)$ contributions from $\nu_e$, $\nu_\mu$, and $\nu_\tau$ with a broad range of energies as described above. In what follows we outline how to formalize the problem of atmospheric neutrino upscattering $\nu A \rightarrow N X$, with $A$ a SM nucleus and $X$ some SM final state particles.

For our upscattering formalism, we begin with the incoming flux of atmospheric neutrinos. The flux of these neutrinos is sensitive to the neutrino energy and the zenith angle relative to the neutrino entry point.\footnote{In general one could also consider neutrino fluxes that depend on latitude and longitude, however the Honda fluxes \cite{Honda:2015fha} are computed only at a few select locations and so we treat cosmic ray production identically at all locations on the Earth's surface.} For flavor dependent couplings we include neutrino oscillations, which are affected by the matter profile between the entry position, $\vb{W}$, and the interaction point $\vb{X}$. The result is an angle and energy dependent neutrino intensity $\mathcal{I}_{\nu_{\alpha}}(E_{\nu},\Omega_\nu,\vb{X})$ that depends on the neutrino flavor $\alpha$, and the position inside the Earth, $\vb{X}$. This intensity can be related to the standard atmospheric neutrino intensity of flavor $\beta$, $I_{\nu_\beta}$, (at the surface) via 
\begin{equation}
    \mathcal{I}_{\nu_{\alpha}}(E_{\nu},\Omega_\nu,\vb{X}) = P_{\alpha\beta}(\vb{X},E_\nu,\Omega_\nu) I_{\nu_{\beta}}(E_\nu, \cos\phi_{\rm zen})
\end{equation}
where a sum over $\beta$ is implied, and $\cos\phi_{\rm zen}$ is chosen so that the neutrino points from $\vb{W}$ to $\vb{X}$ (see \cref{cartoon}). Neutrino oscillation probabilities, denoted by  $P_{\alpha\beta}$ for $\nu_\beta\rightarrow \nu_\alpha$, depend on the position inside the Earth and the angle of incidence since these two parameters determine the neutrino's path through the Earth. The zenith angle at which the neutrino is produced also depends on both $\vb{X}$ and $\Omega_\nu$. The neutrino oscillations must be computed separately for each angle, energy, and point inside the Earth. In what follows we take recent best fit values from the NuFit collaboration~\cite{Esteban:2020cvm}: $\Delta m_{21}^2= 7.42\times 10^{-5}~{\rm eV}^2$, $\Delta m_{31}^2= 2.52\times 10^{-3}~{\rm eV}^2$, $\theta_{12}=33.44\deg$, $\theta_{13}=8.57\deg$, $\theta_{23}=49.2\deg$, and $\delta_{\rm CP} = 197\deg$. The effect of varying neutrino oscillation parameters within their allowed range of values produces only a small effect. This is because the flux of HNLs arriving at the detector depends on the volume-averaged oscillation probabilities weighted by the broadband atmospheric flux. The result therefore samples a wide range of $L/E_\nu$ and is relatively insensitive to e.g.\ $\delta_{\rm CP}$ and the mass mixing hierarchy. 

While most of the atmospheric neutrino flux has energies of 100s of MeV, the flux extends to 10s of TeV. Momentum transfers can then be much larger than the scale of nuclear coherence $Q_{\rm coh} \sim 100$ MeV such that scattering will not be entirely coherent. Instead, the cross section be composed of a coherent and incoherent piece $\dd\sigma = \dd\sigma_{\rm coh} + \dd\sigma_{{\rm in.} }$ where $\dd\sigma_{\rm coh}\sim O(Z^2,Q_w^2)$ and $\dd\sigma_{{\rm in.} }\sim O(Z,Q_w)$ with $Z$ an $Q_w$ the electric and weak charge of the nucleus. The former is relevant for the dipole portal and the latter is relevant for the mass mixing portal. The coherent contribution can be reliably treated working in the infinite mass limit because nuclear form factors ensure that $|Q|\sim 0.3~{\rm GeV}$ whereas nuclei are generically heavy $M_A \sim 30~{\rm GeV}$ such that nuclear recoil energies are small $T_R\sim Q^2/2 M_A\ll E_\nu$.

In general the upscattering cross section will depend on both the scattering angle and HNL energy $\dd^2 \sigma /\dd \cos\Theta \dd E_N$. We are interested in angles, $\Theta$, such that the resulting HNL is directed toward the detector. When considering coherent scattering, the nucleus' recoil can be neglected such that 
\begin{equation}
    \qty[ \frac{\dd^2 \sigma}{\dd \cos\Theta \dd E_N}]_{\rm coh} = \frac{\dd \sigma}{\dd \cos\Theta} \delta (E_N - E_{\nu})
\end{equation}
For elastic scattering on {\em free} nucleons, recoil effects must be included and the delta function instead relates $E_N$ as a function of both $E_\nu$ and $\cos\Theta$. In this work we include the contribution from coherent scattering and incoherent scattering on the constituent nucleons. We model the incoherent contribution as if the scattering took place on free nucleons and neglect detailed nuclear effects. In models where the upscattering is dominated by coherent scattering (e.g.\ the dipole portal) the nuclear uncertainties are drastically reduced. For the mass-mixing portal we find that incoherent scattering provides an $O(1)$ contribution to the total rate, and treat the theory uncertainty from nuclear effects conservatively (see \cref{subfig-U}).
%
%
%

The HNL created in this interaction is unstable, with a decay length of 
\begin{equation}
    \lambda = \gamma \beta \tau~,
    \label{eq:Gen_Decay_Length}
\end{equation}
where $\tau$ is the characteristic decay time in the rest frame of the HNL. Given the decay length, $\lambda$, the probability for an HNL produced at a location $\vb{X}$  directed toward our detector located at $\vb{Y}$ to decay visibly within the fiducial volume, of length $\Ldet$ and area $\Adet$ is 
\begin{equation}
    \Pvis = B_{\rm vis}\times \exp\bigg(\,-\frac{|\textbf{X} -\textbf{Y}|}{\lambda}\bigg)\, \bigg(\,1 - \e^{-\Ldet/\lambda}\, \bigg)\,~,
    \label{eq:P_Dec}
\end{equation}
where $B_{\rm vis}=\Gamma_{N\rightarrow \rm vis}/ \Gamma_N$ is the branching ratio to visible SM decay products (an experiment and search strategy dependent quantity). The probability of an HNL being directed towards the detector is proportional to the solid angle of the detector as seen from the point of emission, $\Omega_{\rm det} \sim \Adet/|\textbf{X}-\textbf{Y}|^2$. We note that in the limit $\lambda \gg \ell$ we may approximate $1-\e^{-\ell/\lambda}\approx \ell/\lambda$ such that the overall rate scales as the volume of the detector $ V=A_\perp \ell $. For $\lambda \sim \ell$ the rate will depend somewhat on the geometry of the detector and so the ceiling of our constraints will be modified by an $O(1)$ number; because this region is already ruled out by complementary search strategies this is not important, except perhaps for electron-only coupled dipole portals (see \cref{subfig-e}). Putting everything together the event rate for HNLs produced by a $\nu_{\alpha}$ neutrino portal (with $\alpha \in \{e,\mu,\tau\}$) to decay visibly inside the fiducial volume of a detector is given by
\begin{widetext}
\begin{equation}\label{master-eq}
        \dv{R_{N\rightarrow {\rm vis}} }{E_N} =   \int_{\oplus} \dd^3X  
   \qty[\int \dd E_\nu \int \dd\Omega_\nu     \sum_i n_i(\vb{X})\frac{1}{4\pi|\textbf{X} - \textbf{Y}|^2} \mathcal{I}_{\nu_{\alpha}}(E_{\nu},\Omega_\nu,\vb{X})    \frac{\dd^2 \sigma_i}{\dd \cos \Theta \dd E_N}] \Adet \Pvis(E_N, |\vb{X}-\vb{Y}|)~.
\end{equation}
\end{widetext}
 The term in square brackets is the differential flux of HNLs per unit volume produced at location $\vb{X}$ and the factor $P_{\rm vis}$ weights the spectrum by the probability of decay within the detector. The event spectrum for a given visible decay product can be found by folding the differential rate $\dd R/\dd E_N$ computed using \cref{master-eq} with the spectrum of daughter particles produced in the lab frame by an HNL with energy $E_N$.

\Cref{master-eq} cannot be calculated analytically for simple matter profiles due to the complex dependence of the oscillated neutrino intensity as a function of $\vb{X}$. Even without oscillations, a realistic density and composition profile of the Earth demands a numerical treatment. We have developed a purpose built Monte Carlo program capable of solving \cref{master-eq} efficiently using conditional importance sampling. The details of our implementation are discussed in \cref{Monte-Carlo}, however we briefly sketch the procedure here. First, we generate an ensemble of neutrino energies by importance sampling an approximate atmospheric neutrino flux curve. For each neutrino energy we calculate the maximum HNL decay length, which corresponds to the case when $E_N=E_\nu$ such that $\lambda_{\rm max}=\lambda(E_N=E_\nu)$. We then sample a position inside the Earth, $\vb{X}$, from an exponential distribution defined relative to the detector. At each point $\vb{X}$ the density and composition of the Earth is computed. For each production mechanism (e.g. coherent v.s. incoherent scattering off $^{56}$Fe and $^{16}$O), we generate a random initial neutrino angle (defined relative to $\vb{Y}-\vb{X}$) using a non-uniform sampling that accounts for correlations induced by the differential cross section\footnote{For highly forward scattering these correlations can make certain numerical methods (e.g.\ \texttt{VEGAS}) highly inefficient \cite{Eby:2019mgs,Paddy-disc}. By ``working backwards'' from the detector to the source of neutrinos we efficiently account for these correlations. } $\dd \sigma_i/ \dd \cos\Theta$. Given the incident neutrino angle, we then propagate backwards to the point on the Earth's surface where the neutrino would have originated, $\vb{W}$, and we calculate the zenith angle relative to the Earth's tangent at that point. The neutrino intensity is then calculated using \texttt{NuFlux}~\cite{NuFlux:2022} with $E_\nu$ taken from the first step, and at the required zenith angle to propagate from $\vb{W}$ to $\vb{X}$. All events are saved in an event record, with appropriate weights accounting for the various terms in \cref{master-eq}. Finally, in a post-processing stage we calculate the relative weights for the various neutrino flavors at the location $\vb{X}$ by numerically solving the Schr\"odinger equation along the line segment connecting the upscattering location $\vb{X}$ to the neutrino point of origin at the Earth's surface. This is done self-consistently using the same density and composition profile as was used to generate the upscattering events.

We now specialize our discussion to the relevant neutrino portals discussed herein. 

\subsection{Dipole Portal}
For the dipole portal, the visible decay signal is $N\rightarrow \gamma\nu$ with a branching ratio of ${\rm BR} =1$. The analysis is consequently straightforward. The decay length of the HNL is given by
\begin{equation}
    \begin{split}
        \lambda &= \frac{4 \pi}{d^2 m_N^3} \gamma \beta ~,
    \end{split}
    \label{eq:Decay_Length}
\end{equation}
which is comparable to the size of the Earth for much of the parameter space of interest. 

\begin{figure}[t]
    \includegraphics[width=.95\linewidth]{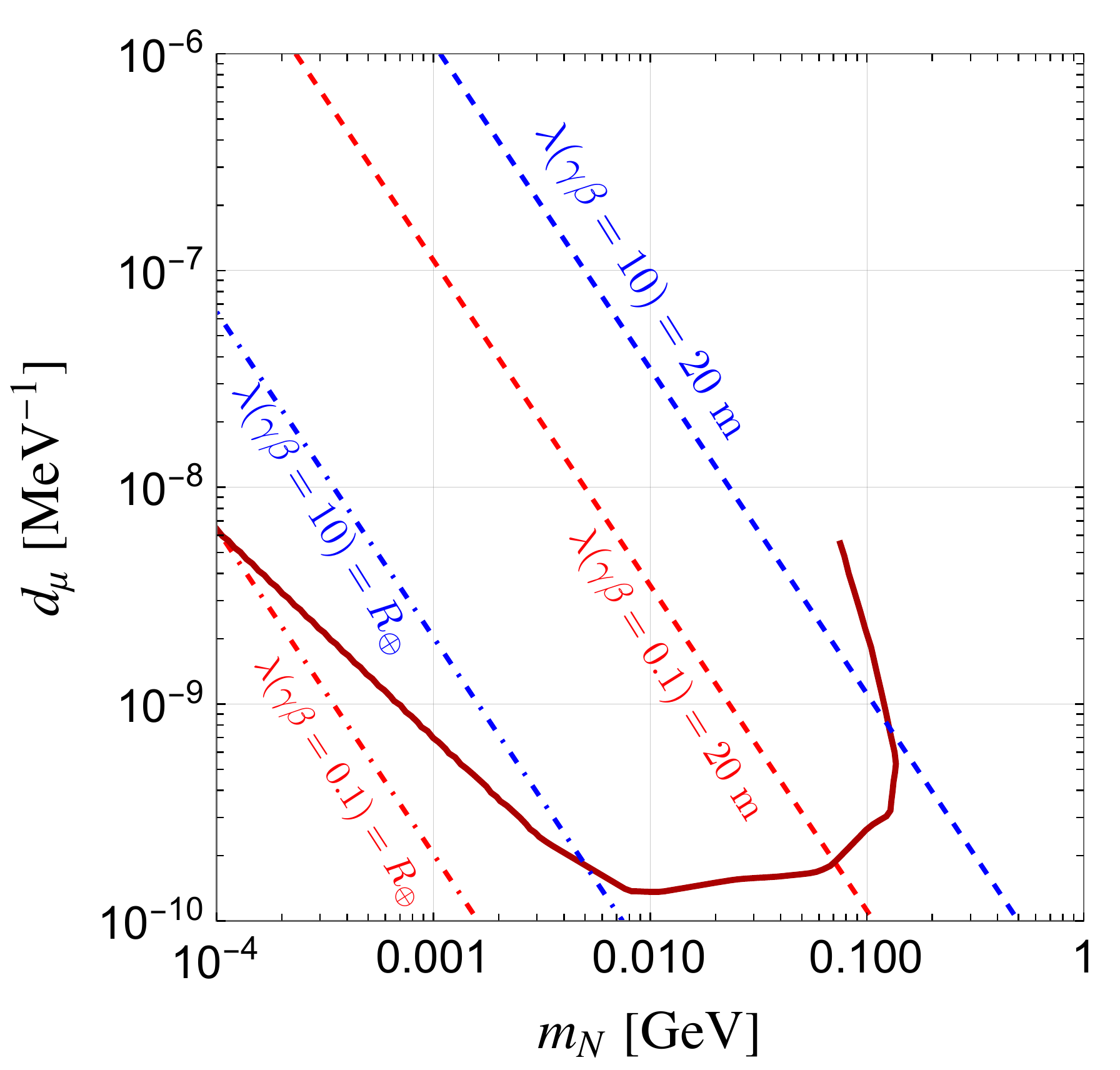}
    \caption{Combined exclusion contours ignoring oscillations, along with relevant decay lengths. This plot is for the flavor independent dipole coupling ($d_e = d_{\mu} = d_{\tau}$). \label{qualplot}}
\end{figure}

For a neutrino dipole portal, coherent scattering on nuclei is the dominant upscattering process for all energies. The upscattering cross section assumes a simple form when one  works in the large mass limit and drops all mass-suppressed effects including nuclear recoil corrections, and contributions from a nuclear magnetic dipole moment. This is a valid approximation because the nucleus' charge form factor ensures that $\vb{Q}\ll 300$ MeV, such that recoil corrections are always small. The resulting differential cross section is of the form 
\begin{equation}
    \dv{\sigma_{\rm coh}}{\vb{Q}^2} \sim   \frac{4Z^2 \alpha d^2}{\vb{Q}^2} |F(\vb{Q}^2)|^2 \times \qty(1- \frac{m_N^2}{4 E_\nu^2} - \frac{m_N^4}{E_\nu^2\vb{Q}^2})~.
    \label{eq:Coh_Cross_Sec}
\end{equation}
The angular dependence $\dd \sigma/ \dd \cos\Theta$ can be obtained by a simple change of variables. The charge form factor of each nucleus is modelled as a Helm form factor with parameters fitted to the tabulated two parameter Fermi distributions from \cite{DeVries:1987atn} (see \cref{form-factors} for more discussion).

Although it is subdominant to the coherent contribution, we also include an incoherent sum over nucleons in our model of upscattering. This cross section is given by $\dd \sigma_A = Z \dd \sigma_p+ (A-Z)\dd \sigma_n$ with the proton and neutron cross sections parameterized in terms of standard Dirac and Pauli form factors.

\begin{figure}
    \includegraphics[width = 1 \linewidth]{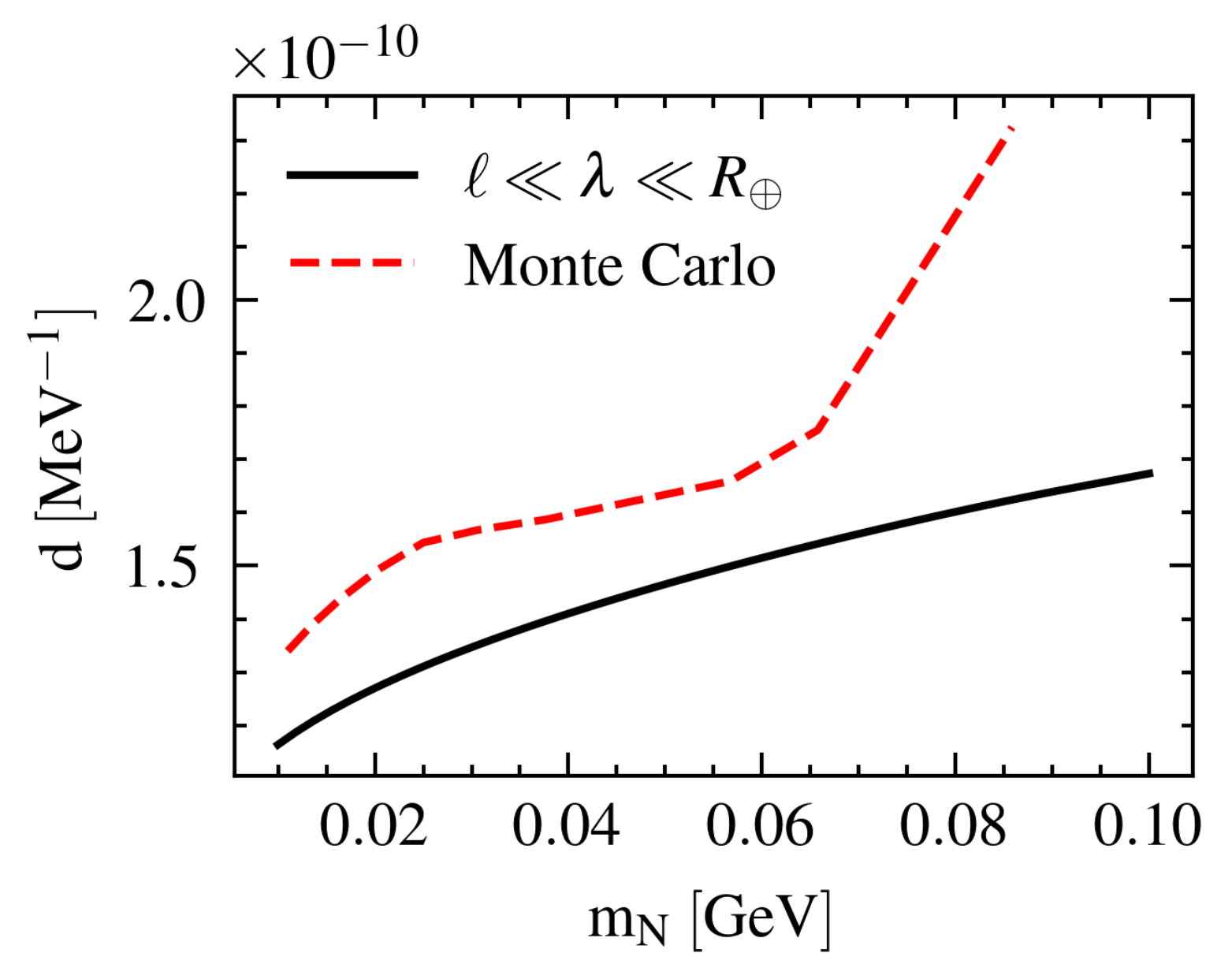}
    \caption{Flavor independent curves corresponding to 197 HNL events at Super-Kamiokande with 5326 days of data using the approximation from \cref{floor_approximation} and the Monte Carlo simulation. In the approximation, we consider the Earth as composed entirely of silicon with a density of 5 $\mathrm{g/cm^3}$.}
    \label{Approx_vs_Monte_Carlo}
\end{figure}

We can see in \cref{money-plot} that our dipole coupling bounds have a relatively flat region in d when $m_N$ is between 0.01 and 0.08 GeV. In \cref{qualplot}, we see that this flat region corresponds to decay lengths satisfying the hierarchy $\ell \ll \lambda \ll R_{\oplus}$. This bound can be estimated through a relatively simple approximation: treat the Earth as being of a constant density composed of a single element, consider the neutrino flux as isotropic, ignore angular dependence on the cross section, consider elastic scattering such that $E_{\nu} = E_N$, and set all terms of $O(\ell / \lambda)$ and $O(\lambda / R_{\oplus})$ to zero. Within this approximation, we find
\begin{equation}
    \frac{\dd R_{N \rightarrow {\rm vis}}}{\dd E_N} = \frac{n V_{\rm det}}{2} I_{\nu}(E_N) \sigma(E_N) f_{\rm vis}(E_N),
\end{equation} 
where $f_{\rm vis}$ is the fraction of HNL decays in our detector that are in the visible energy range. Here, the only depenedence on the transition dipole moment appears in the cross section as a $d^2$ (the rate is independent of the decay length for $\ell \ll \lambda \ll R_{\oplus}$). We can define $\Tilde{\sigma} = \sigma / d^2$, and then our estimate for the floor of our constraint is
\begin{equation}
    d(m_N) = \sqrt{ \frac{2 R^{\rm exp}_{N \rightarrow {\rm vis}}}{n V_{\rm det} \int \dd E_N I_{\nu}(E_N) \Tilde{\sigma}(E_N) f_{\rm vis}(E_N)}}~,
    \label{floor_approximation}
\end{equation} 
where $d(m_N)$ is an estimate for the ``floor'' of our constraint as a function of $m_N$ and $R^{\rm exp}_{N \rightarrow {\rm vis}}$ is the rate of visible energy deposition that can be excluded by the experiment under consideration. In \cref{Approx_vs_Monte_Carlo}, we see that this approximation closely matches the true bounds that we get for the full Monte Carlo, meaning that this approximation can be used to see how the lower bound will scale with exposure time, volume, decreased background, etc. Since this approximation ignores decay length, it should also be valid for setting approximate bounds on Majorana HNLs, which have half the decay length of Dirac HNLs

\subsection{Mass Mixing Portal} 
The mass mixing portal is more complicated phenomenologically because production cross sections rise with energy, and for $m_N \gtrsim m_\pi$ many new hadronic decay channels open. We have included many of these details in our simulation, however \emph{a posteriori} it is clear that searches relying on terrestrial upscattering are only competitive with existing constraints for masses below the pion threshold. We therefore focus our discussion on the case of the decay channel $N \rightarrow e^+e^- \nu$ which is the only visible decay mode for $m_N \leq 135$ MeV. 

The decay of an HNL to an $e^+e^-$ pair depends on the flavor structure of the mass mixing portal and the flavor of the invisible SM neutrino. These effects introduce an $O(1)$ prefactor that depends on the final state  which can be found in ~\cite{Gorbunov:2007ak,Bondarenko:2018ptm,Coloma:2020lgy}, however the dominant effect is the muon decay like formula for the partial width
\begin{equation}
    \Gamma_{e^+e^- \nu} = \frac{G_F^2 m_N^5 |U|^2}{192 \pi^3} \times O(1) ~. 
\end{equation}
The result is that HNL decay lengths are extremely long for low masses, and can easily exceed the radius of the Earth by orders of magnitude. In this regime terrestrial upscattering offers substantial benefits over traditionally laboratory based searches, and can offer leading sensitivity on $|U_{\tau N}|^2$. 

HNLs can always decay to three neutrinos, $N\rightarrow \nu \nu\nu$ for any non-zero mixing angle. The result is a branching ratio that is $O(10\%)$ for $N\rightarrow \nu e^+e^-$ for all HNL masses below the pion threshold. We include this effect in our simulations computing the full decay length and taking $B_{\rm vis} = \Gamma_{e^+e^-\nu}/\Gamma$. 

HNL upscattering proceeds via the weak neutral current and for the relatively low HNL masses that we focus on here all of coherent (i.e.\ CEvNS), quasi-elastic, and deep inelastic scattering contribute to the upscattering yield. We find that for regions of parameter space where atmospheric upscattering is competitive that the scattering mechanisms are dominantly coherent and incoherent scattering on nucleons, with deep inelastic events contributing only a few percent to the total flux. 

The coherent contribution is relatively insensitive to the nuclear species and can be calculated from first principles. We model the weak nuclear form factor by setting it equal to the charge nuclear form factor. Incoherent scattering on nuclei is modelled as described above for the dipole portal. This neglects all effects of nuclear structure and we therefore expect a sizable theoretical error from our modelling. Unlike the dipole portal case, we find that incoherent scattering makes up roughly two-thirds the total upscattered flux. Owing to its relative importance, we have included ``error bands'' in \cref{money-plot} in which the incoherent scattering cross section has been doubled, and halved respectively; we believe this to be a conservative overestimate of the theoretical uncertainty.

\subsection{Decays inside the Detector \label{decays} } 
In the presentation above we have outlined how to calculate the flux of unstable particles arriving at a given large volume detector. This flux is not directly visible, and the bona fide observable is the energy and angular distribution of an HNL's visible daughter particles. For illustration, we discuss the case of a dipole portal decay $N \rightarrow \nu \gamma$ in detail below. The case of a three-body decay, as in $N\rightarrow \nu e^+ e^-$ is qualitatively similar, but slightly more involved due to the three body final state. The details of the decay distribution do not substantially impact our rate-only estimate, although their details may be relevant for future searches that we outline in \cref{dedicated}. 

In the case of the dipole portal, when the HNL decays, it decays into a photon and neutrino. The angular distribution of a dipole-mediated decay in the HNL rest frame depends on the level of $CP$ violation \cite{Balantekin:2018ukw,Balantekin:2018azf,Kayser:2018mot,Balaji:2019fxd,Sierra:2021say}, with $\dd \Gamma/\dd\cos\zeta' \sim 1+\alpha \cos\zeta'$ and $\alpha\in[-1,1]$ and $\zeta'$ the angle between the photon and the HNL polarization. For simplicity, we take $\alpha=0$ such that the decays are isotropic. Our sensitivity is only mildly sensitive to this choice; $\alpha>0$ leads to a somewhat harder photon spectrum in the lab frame, while $\alpha<0$ leads to a somewhat softer spectrum (see e.g.\ related discussion in \cite{Plestid:2020vqf}). In the rest frame, $E_{\nu,\rm rest}$ = $E_{\gamma, \rm rest}$ = $E_N/2$, and this leads to the following lab frame kinematic variables
\begin{align}
    \tan(\zeta_{\rm lab}) &=  \frac{m_N}{E_N} \frac{\sin(\zeta')}{\cos(\zeta') + \sqrt{1 - m_N^2/E_N^2}}~,\\
    E_{\gamma, \rm lab} &= \frac{E_N}{2} \bigg(\, 1+\cos (\zeta') \sqrt{1 - \frac{m_N^2}{E_N^2}} \bigg)~, 
\end{align}
A flat (i.e.\ isotropic) distribution in $\cos\zeta'$ results in a ``box distribution'' for $E_{\rm \gamma, lab}$ ranging between $[E_{\rm \gamma, lab}^{(-)},E_{\rm \gamma, lab}^{(+)}]$ where $E_{\rm \gamma, lab}^{(\pm)}=\tfrac12 E_N( 1\pm  \sqrt{1 - m_N^2/E_N^2})$.
Knowing the initial momentum of the HNL, we can sample $\cos\zeta'$ uniformly between $[-1,1]$ and generate a random sample of angles of the detected photon relative to the horizon at the detector $\phi_{\rm det}$. 

\section{Super-Kamiokande constraints \label{results} } 
We now turn to our analysis of public data from Super-Kamiokande, which when coupled with our Monte Carlo simulation, allows us to set new limits on neutrino portal couplings. SK is a large volume ($22.5\times 10^3 ~{\rm  m}^3$ fiducial volume) Cherenkov detector whose primary background is the scattering of atmospheric neutrinos passing through the detector. It is well suited to search for through-going HNL decays and has a large statistical sample of atmospheric neutrino events which can be used to set limits on the rate of visible HNL decay \cite{Super-Kamiokande:2014ndf,Super-Kamiokande:2019gzr}. 

The SK collaboration classifies events as sub-GeV ($30 ~{\rm MeV}<E_{\rm vis} < 1.33 {\rm~ GeV}$) and multi-GeV ($E_{\rm vis} > 1.33~{\rm GeV}$) with sub-classifications for each event type. In the sub-GeV sample events are classified as e-like, $\mu$-like or $\pi^0$-like, single-ring or two-ring, and 0 decay-e, 1 decay-e, or 2 decay-e. The decay-e classification is meant to capture Michel-electrons from muon decay, while the particle identification (PID) is based on characteristic Cherenkov ring patterns of each particle. The multi-GeV sample is split into partially contained and fully contained, the former applying exclusively to muon events. In the fully contained sample events are classified as single-ring or multi-ring and are then further sub-divided as $\nu_e$-like, $\bar{\nu}_e$-like, or $\mu$-like. The $\nu_e$-like vs $\bar{\nu}_e$-like samples are defined by a cut on the number of decay-e events which ultimately stem from a $\nu_e n \rightarrow e^- \pi^+ n$ interaction with subsequent pion decay at rest, followed by muon decay \cite{Super-Kamiokande:2017yvm}. Not all $\nu_e$ interactions produce a $\pi^+$ and so there is substantial cross-contamination between the two samples; by way of contrast the $\mu$-like sample is relatively pure. 

In what follows we describe a simple rate-only analysis based on the published results in \cite{Super-Kamiokande:2017yvm}. For each model we focus on a the relevant experimental signature and use the experimental collaboration's Monte Carlo prediction as the expected Poisson mean of the event sample. Given their observed data, we then set limits at the 95\% confidence level on the number of allowed events in the energy range as defined by the experiment. We consider both systematics and statistically limited searches with a conservative estimate of a $5\%$ systematic uncertainty on the collaboration's Monte Carlo prediction for their sub-GeV sample of 0-decay-e events. 
\subsection{Dipole Portal} 
The dipole portal's only signature is a single photon which will be classified as an e-like 0 decay-e signature in the sub-GeV analysis and as a fully-contained $\bar{\nu}_e$-like event in the multi-GeV sample. The energy distribution of photons is broad for all HNL masses but its precise shape depends on both $m_N$ and $d$. The multi-GeV and sub-GeV samples therefore provide complementary tools with which to probe the HNL parameter space. 

\begin{figure}
    \includegraphics[width=1\linewidth]{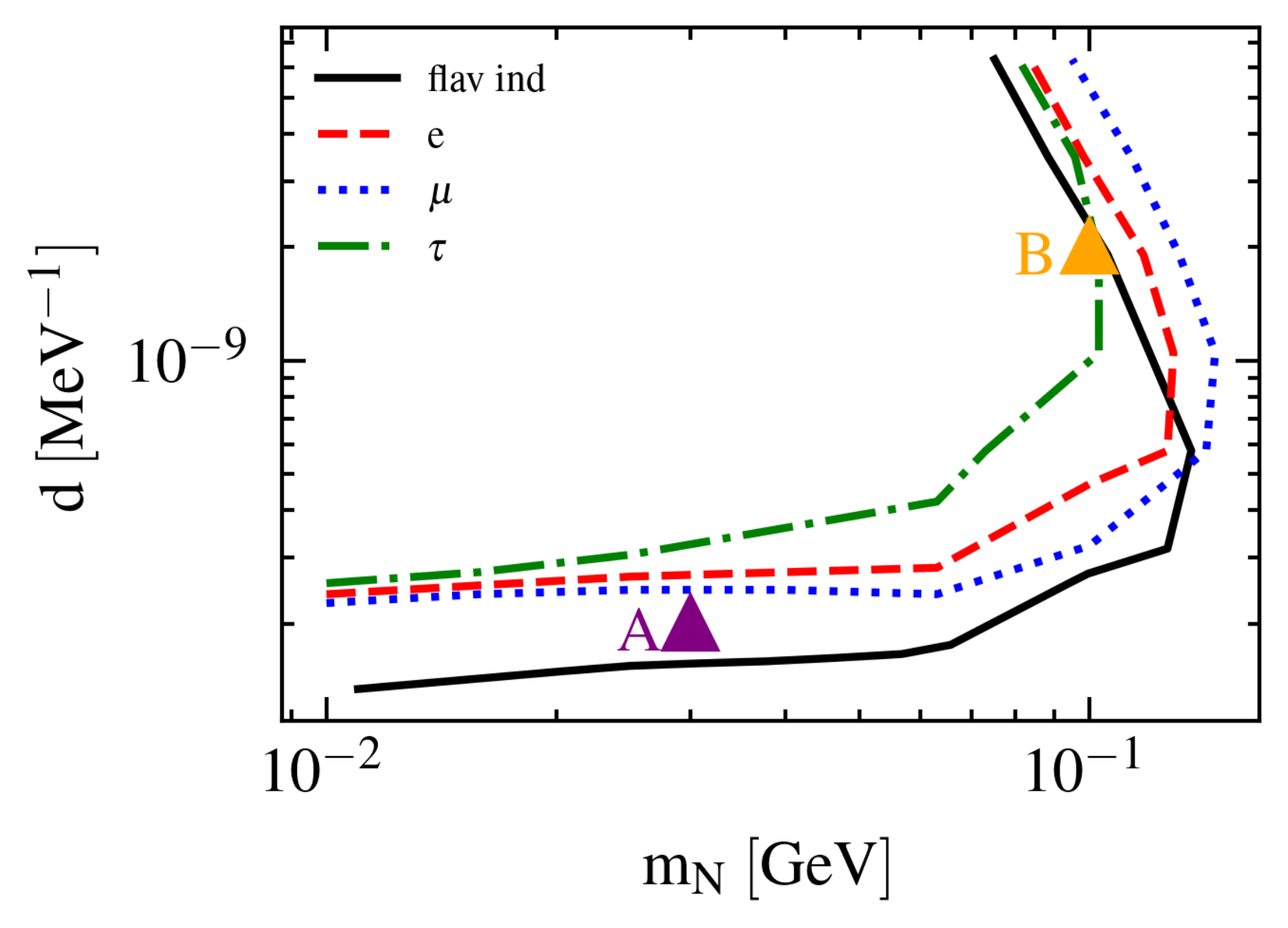}
        \caption{Exclusion contours assuming a statistically limited search (i.e.\ $\sigma_{\rm sys} =0$)  from \cref{money-plot} (a), (b), and (c) and the equivalent constraint for a flavor independent dipole portal. Our constraints are only moderately sensitive to neutrino flavor due to substantial oscillations within the interior of the Earth. Point A is representative of parameter space that is dominated by the sub-GeV sample at Super-K. Point B is representative of a point in parameter space that is dominated by the multi-GeV sample. We discuss this further in \cref{dedicated} and show distributions for Point A in \cref{Super_K_angles_lower,Super_K_energies_lower} and Point B in \cref{Super_K_angles_upper,Super_K_energies_upper}. } \label{Super_K_Plot}
\end{figure}

We set limits by taking the union of the excluded regions from the multi-GeV and sub-GeV analyses separately and these are shown in \cref{Super_K_Plot}. We note that the constraints cross around $d\simeq 5 \times 10^{-9}~{\rm MeV}^{-1}$, with the multi-GeV search dominating for larger $d$ and the sub-GeV dominating below. We consider $d_{e} = d_{\mu} = d_{\tau} = d$ (flavor independent), and flavor dependent couplings accounting for neutrino oscillations in each case. Based on Table II of \cite{Super-Kamiokande:2017yvm}, we assume a Poisson mean from the collaboration's Monte Carlo simulation of $\mu_{\rm MC}=10266$ sub-GeV 0 decay-e events while the observed count is $N_{\rm obs} = 10294$. 

It is tempting, given the close agreement between Monte Carlo and observation to infer that systematic uncertainties on the Monte Carlo prediction are fully under control, however the Super-Kamiokande Monte Carlo is tuned to their data to self-consistently determine, e.g.\ the atmospheric flux normalization. In the presence of new physics this tuning could be compromised and so it is important to estimate a systematic uncertainty on an experiment such as Super-Kamiokande. First, note that while the overall normalization of the atmospheric flux is poorly constrained, the $\nu_e: \nu_\mu$ ratio is known to within a few-percent \cite{Super-Kamiokande:2017yvm,Honda:2015fha}. Therefore the flux normalization at Super-Kamiokande can be fixed using muon-exclusive sub-samples, and the electron flux can be subsequently inferred. Second, it is worth noting that the sub-GeV 0-decay e bin of the Super-Kamiokande data set is relatively insensitive to neutrino oscillations, and its background modelling is therefore reasonably robust. Quantitatively, one can compare the predicted flux with and without oscillations from Fig.\ 14.4  of \cite{Zyla:2020zbs}; the flux changes by only 2.7\%. We therefore conclude that a 5\% systematic uncertainty can be conservatively applied to the Super-Kamiokande Monte Carlo prediction of the 0-decay e sub-GeV background from atmospheric $\nu_e$ scattering.

For finding constraints, we take the statistical uncertainty as $\sigma_{\rm{stat}} = \sqrt{\mu_{MC}+\mu_{HNL}}$. We take a conservative upper bound on the systematic uncertainty at $\sigma_{\rm{sys}} = 0.05  \mu_{MC}$. We then solve $P(x \leq N_{\rm obs} | \mu, \sigma) = 0.05$ where our probability distribution function is $(2 \pi \sigma)^{-1/2} \exp(\frac{(x-\mu)^2}{2 \sigma})$ and $\mu = \mu_{\rm MC} + \mu_{\rm HNL}$ and $\sigma = \sqrt{\sigma_{\rm stat}^2 + \sigma_{\rm sys}^2}$. For $\sigma_{\rm sys} = 0$, we find that $\mu_{\rm HNL} = 197$ is excluded at 95\%-CL; this corresponds to the number of events per 328 kt-yr (corresponding to 5326 live-days at Super-K). For $\sigma_{\rm sys} = 0.05 \mu_{\rm MC}$, we find that our 95\%-CL bound now corresponds to $\mu_{\rm HNL} = 893$. For the multi-GeV analysis we take the $\bar{\nu}_e$-like sample which has  $\mu_{\rm MC}= 2194$ and $N_{\rm obs}=2142$ for a 328 kt-yr exposure. Following the same procedure as above, the $\sigma_{\rm sys} = 0$ 95\%-CL bound is $\mu_{\rm HNL} = 26$ and the $\sigma_{\rm sys} = 0.05 \mu_{\rm MC}$ 95\%-CL bound is $\mu_{\rm HNL} = 145$. It is worth noting that the excluded number of excess events is determined solely by uncertainties in the rate of Standard Model events at Super-K. Uncertainties in our BSM theory only affect how we translate these exclusion bounds into parameter space. We also expect that any uncertainties in our theory are accounted for by our conservative estimate on the Super-K uncertainty.

Using our Monte Carlo integrator, we compute the rate of HNLs passing through and decaying within the detector. The photon spectrum is generated using the lab frame decay distribution of the HNL. For flavor dependent dipole couplings we re-weight the ensemble of Monte Carlo events by adjusting the intensity of neutrinos at the upscattering location according to the oscillation probabilities computed along the line segment connecting the upscatter location to the position on the Earth's surface above which the atmospheric neutrino is produced. Photon detection efficiencies are taken to be unity, $\epsilon_\gamma =1$ which we believe to be reasonable as Super-K can reach 100\% trigger efficiency on events with 4.49 MeV of energy \cite{abe2016solar}, and our photon energies are well above this. The photon spectrum is integrated from $E_\gamma =30$ MeV to $E_\gamma = 1.33 $ GeV for the sub-GeV sample and from $E_\gamma=1.33$ GeV to the highest energy in the Monte Carlo sample for the multi-GeV sample. 


\subsection{Mass Mixing Portal} 

For the mass mixing portal we focus our analysis on $N\rightarrow e^+e^-\nu$ which is the only visible decay mode for $m_N \lesssim 130$ MeV, and contributes for all HNL masses. As we will show the only region in which terrestrial upscattering can compete with fixed target experiments is in the low mass regime and so this suffices for our purposes. 

An $e^+e^-$ pair will appear as highly collimated and result in an electromagnetic shower that is difficult (or impossible) to distinguish from a single electron or single photon. The HNLs are sufficiently boosted such that wide-angle $e^+e^-$ pairs are a non-issue and the decay signature maps onto the same search channels as the single photon analysis. We can therefore take the rate-only exclusions from above an apply them directly to the mass mixing portal. The sub-GeV sample provides the best sensitivity to HNL mass-mixing over the full range of parameter space and we find that new regions of parameter space for $\tau$-coupled HNLs can be probed with existing Super-Kamiokande data. 

For our upscattering simulation we include coherent, quasi-elastic, and deep inelastic scattering channels. We do not include resonance production, nor do we account for nuclear structure (e.g.\ Pauli blocking, giant dipole resonances etc.\ ). We note that in our region of sensitivity, incoherent scattering off of nucleons is the dominant contribution (contributing to around 2/3 of the rate). Coherent scattering contributes to around 25 percent of the total rate, and DIS contributes to less than 10 percent of the rate.
%

\subsection{Scaling with increased sensitivity}
Before moving on to considering future experiments, let us discuss how the constraints described above scale with increased sensitivity. Importantly, the right-most boundary of our exclusions corresponding to an upper bound on  $m_N$ is set by our \emph{sensitivity} rather than by any kinematic thresholds. This is because the flux of atmospheric neutrinos is broad and there is no fundamental limitation on the mass of HNLs which can be produced. As sensitivity improves, either by collecting more data, improving background discrimination, or by leveraging new detector technologies, heavier HNLs can be probed. Furthermore, in the regions of parameter space highlighted in \cref{qualplot}, limits on the dipole coupling scale as $d\sim ({\rm sensitivity})^{1/2}$ which is extremely advantageous relative to the naive scaling of $d\sim ({\rm sensitivity})^{1/4}$ that one would expect in the long-lifetime limit.  Taken together, this suggests that improved sensitivity using atmospheric upscattering as a source of HNLs has a high return on investment.

\section{Search Strategies at Future Experiments \label{dedicated} } %
In this section we discuss how to improve future searches for HNLs. All of the exclusion contours in this work are based exclusively on the simple rate-only estimates of the previous section and the reader may view the following discussion as an outlook towards future improvements. Our Monte Carlo routine can generate kinematic distributions such as the energy and zenith angle of the HNL's decay products, and these can be leveraged to improve signal to background ratios. We also discuss potential improvements in background rejection using different detector technology (e.g.\ liquid scintillator and/or liquid argon time projection chambers instead of a Cherenkov detector), and the impact of a larger fiducial volume in a detector such as Hyper-Kamiokande (HK). 

Let us begin by discussing improvements that can be had by taking into account the energy and angular distributions of the observed photons. For illustration, let us examine two dipole model parameter points which are at the Super-Kamiokande exclusion boundary: the two points marked by triangles in Fig.~\ref{Super_K_Plot}. Recall that these searches exclude a window of couplings, originating from the requirement that decay lengths satisfy: $10$ m $\lesssim \lambda \lesssim R_{\oplus}$. Consequently we will refer to the ``floor'' and ``ceiling'' of the coupling exclusion region.

\begin{figure}[!t]
    \includegraphics[width = 0.97\linewidth]{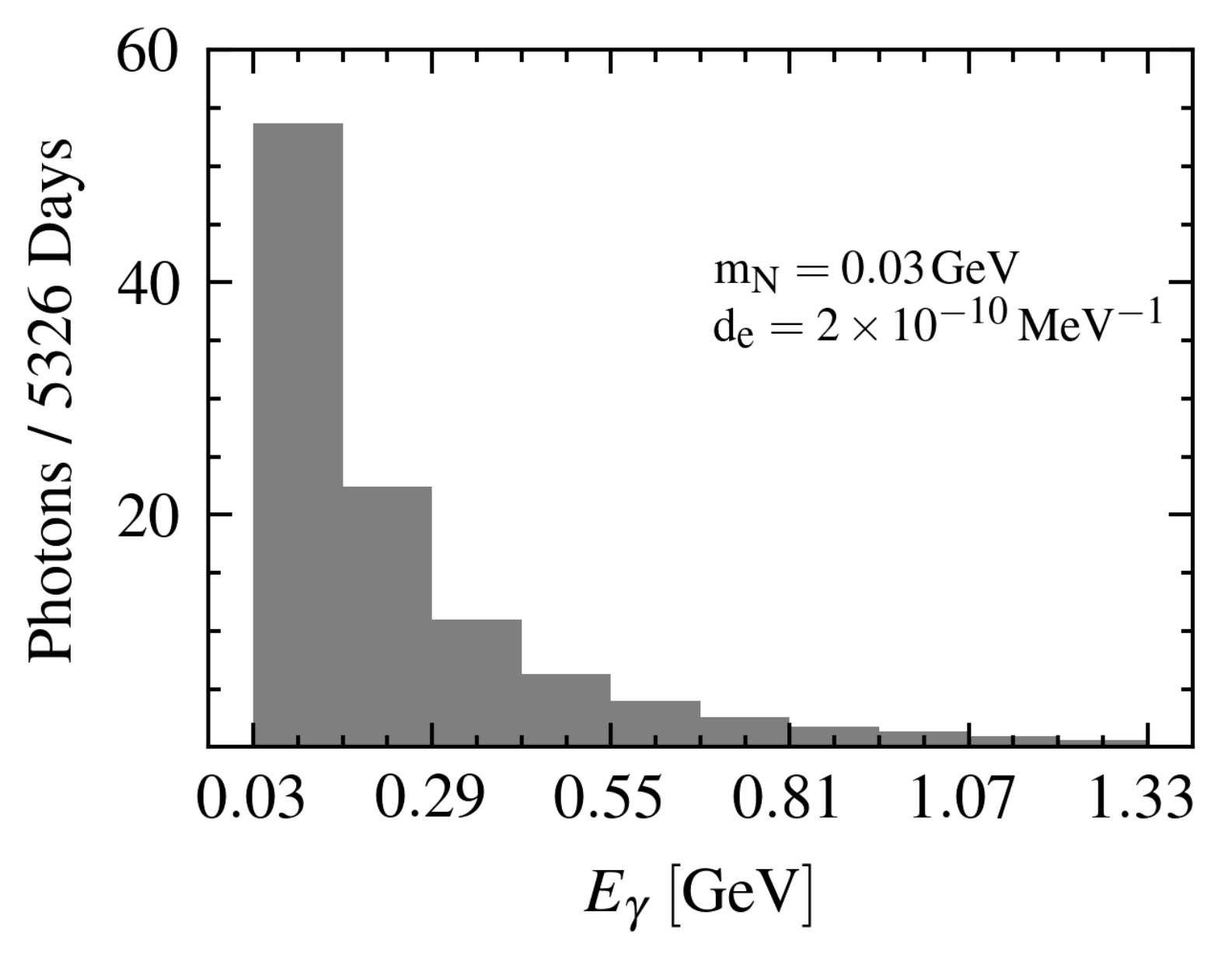}
    \caption{Energy distribution of the detected photons for Point A in \cref{Super_K_Plot}  ($m_N = 0.03$ GeV and $d = 2\times 10^{-10}\, \mathrm{MeV}^{-1}$). At low masses and small dipole couplings, the energy spectrum is IR peaked. The binning in $E_{\gamma}$ is chosen to correspond to the binning in Super-Kamiokande's analysis of the sub-GeV atmospheric neutrino event sample \cite{Super-Kamiokande:2017yvm}. \label{Super_K_energies_lower}}
\end{figure}

Let us first examine the angular and energy distributions of the detected photons from a parameter point on the floor of the exclusion region, $m_N$ = 0.03 GeV and $d_e = 2 \times 10^{-10} \rm{MeV^{-1}}$. We include an example of these distributions on our lower bound for electron-flavor coupling \cref{Super_K_energies_lower}. We see the angular distribution highly favors angles less than $\pi/2$, corresponding to upward going photons in our case. Since the photon direction is highly correlated with the HNL direction, this means most of our signal comes from HNLs produced below the detector. Our decay length is large for these parameters, so there is far more volume available for scattering below the detector than above it. The energy distribution is peaked at lower energies both as a consequence of the atmospheric neutrino flux, which falls off quickly with energy, and because of the skewed distribution of photons from the decay of relativistic HNLs [see discussion near \cref{box-scaling}]. The most probable photon energy in the lab frame is given by one half the HNL's energy. At low HNL masses, almost all of the atmospheric neutrino flux is capable of producing HNLs. For coherent scattering $E_N=E_\nu$, and since the atmospheric neutrino flux falls like a power-law the HNL inherits this feature resulting in lower energy photons.

The angular distribution of observed events at Super-Kamiokande is nearly uniform \cite{Super-Kamiokande:2014ndf}. Therefore, we can choose to only look at upward-going events and cut our background in half while keeping our signal virtually unchanged. We expect that this will extend our lower bound of $d$ by a factor of $2^{1/4}$, as we will need $\sqrt{1/2}$ as many events to reach the same level of uncertainty, and the number of upscattering events goes as $d^2$. 

\begin{figure}[!t]
    \includegraphics[width=0.95\linewidth]{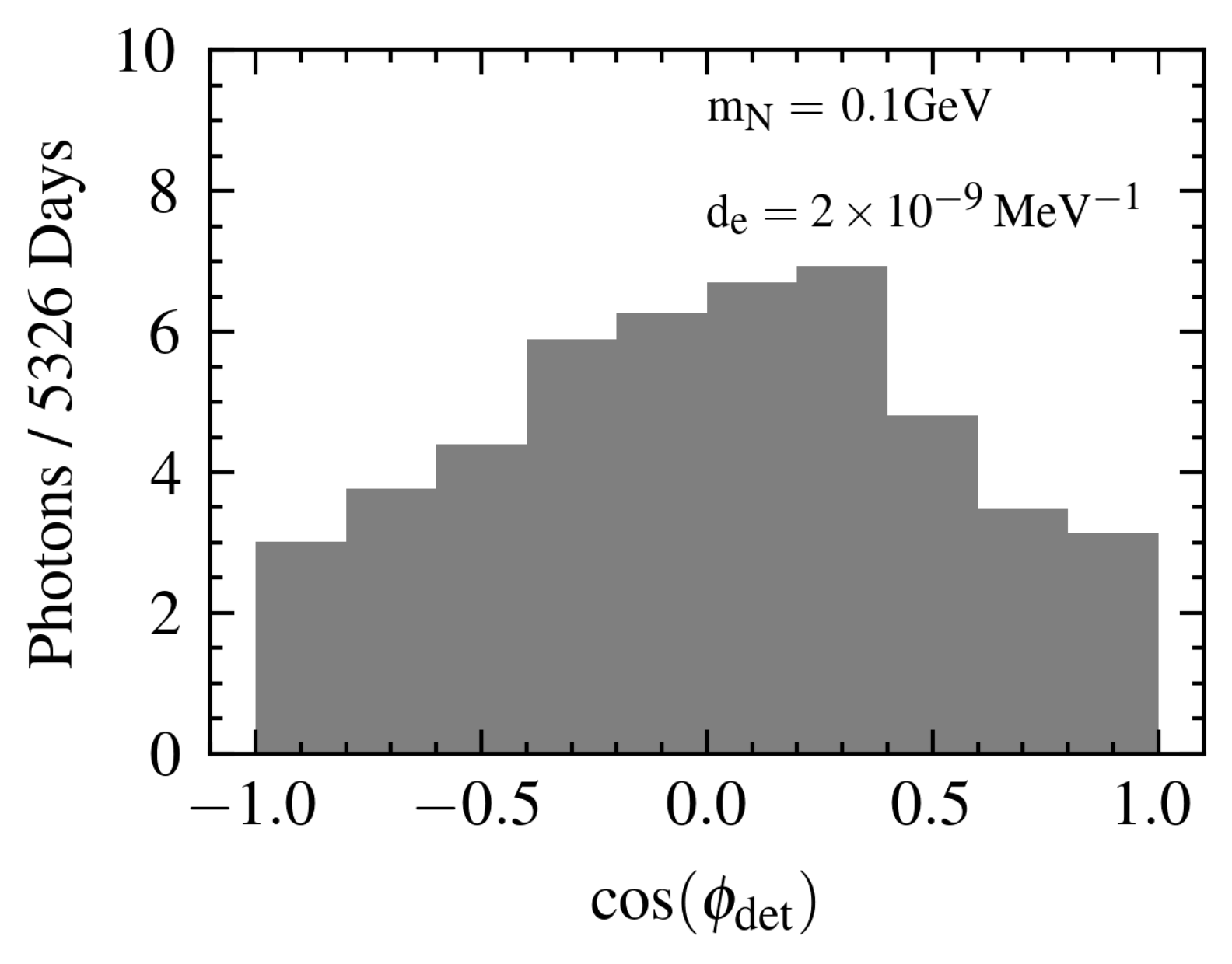}
    \caption{Angular distribution of the detected photons for Point B in \cref{Super_K_Plot}  ($m_N = 0.1$ GeV and $d = 2\times 10^{-9}\, \mathrm{MeV}^{-1}$). The angular spectrum is relatively flat, because at shorter decay lengths, the mountain above Super-Kamiokande contributes an $O(1)$ fraction of the upscattering events. \label{Super_K_angles_upper}}
\end{figure}

\begin{figure}[!t]
    \includegraphics[width=0.95\linewidth]{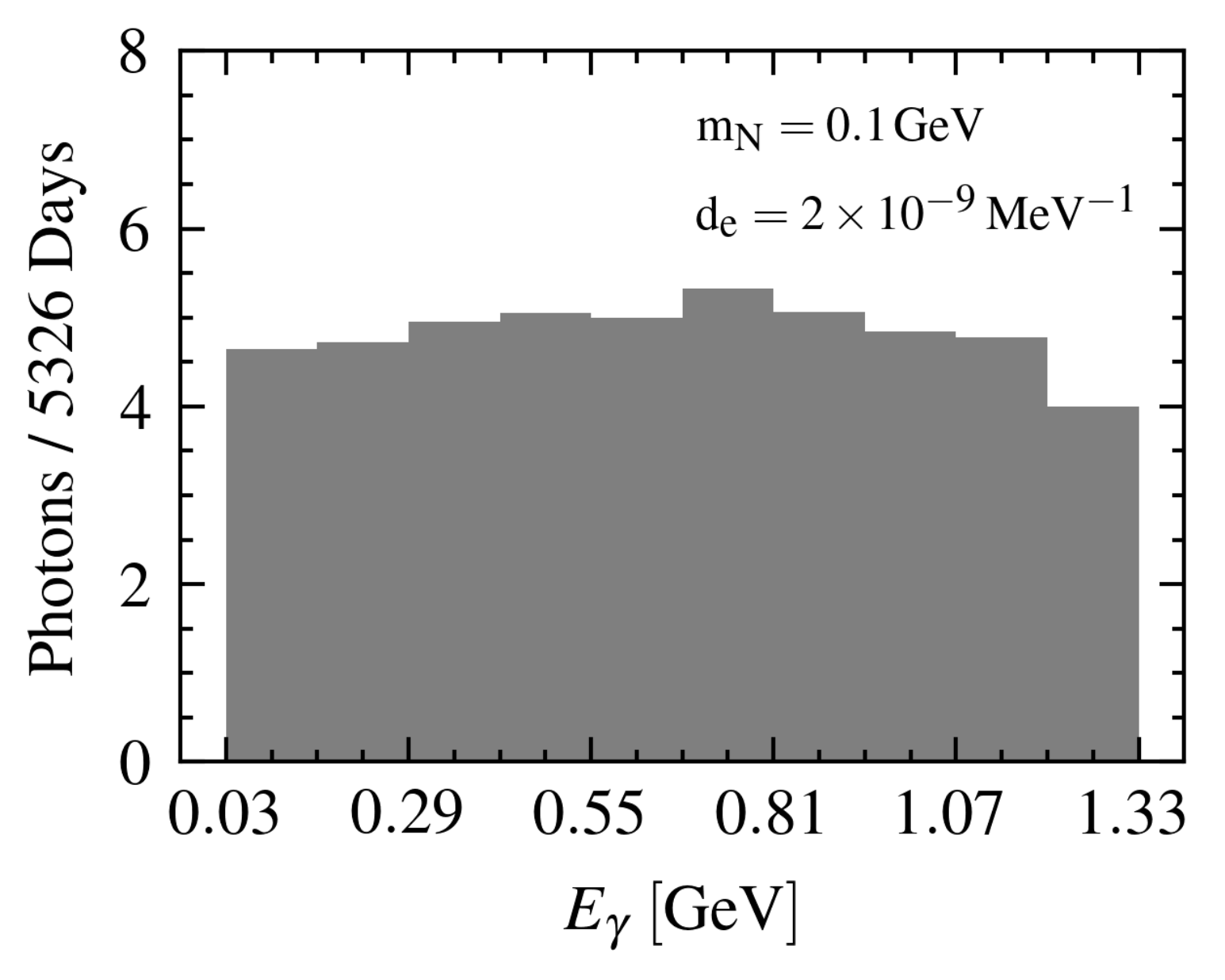}
    \caption{Energy distribution of the detected photons for Point B in \cref{Super_K_Plot}  ($m_N = 0.1$ GeV and $d = 2\times 10^{-9}\, \mathrm{MeV}^{-1}$). At larger masses and stronger dipole couplings, the energy spectrum is relatively flat.  The binning in $E_{\gamma}$ is chosen to correspond to the binning in Super-Kamiokande's analysis of the sub-GeV atmospheric neutrino event sample \cite{Super-Kamiokande:2017yvm}. \label{Super_K_energies_upper}}
\end{figure}

Now let us turn our attention to a parameter point on the ceiling of the excluded region. In this case, the energy and angular distributions are qualitatively different. We show examples of the electron flavor case for $m_N =$ 0.1 GeV and $d_e = 2\times 10^{-9}\, \mathrm{MeV}^{-1}$ (\cref{Super_K_angles_upper} and \cref{Super_K_energies_upper}).
\begin{figure}[!t]
    \includegraphics[height =0.75\linewidth]{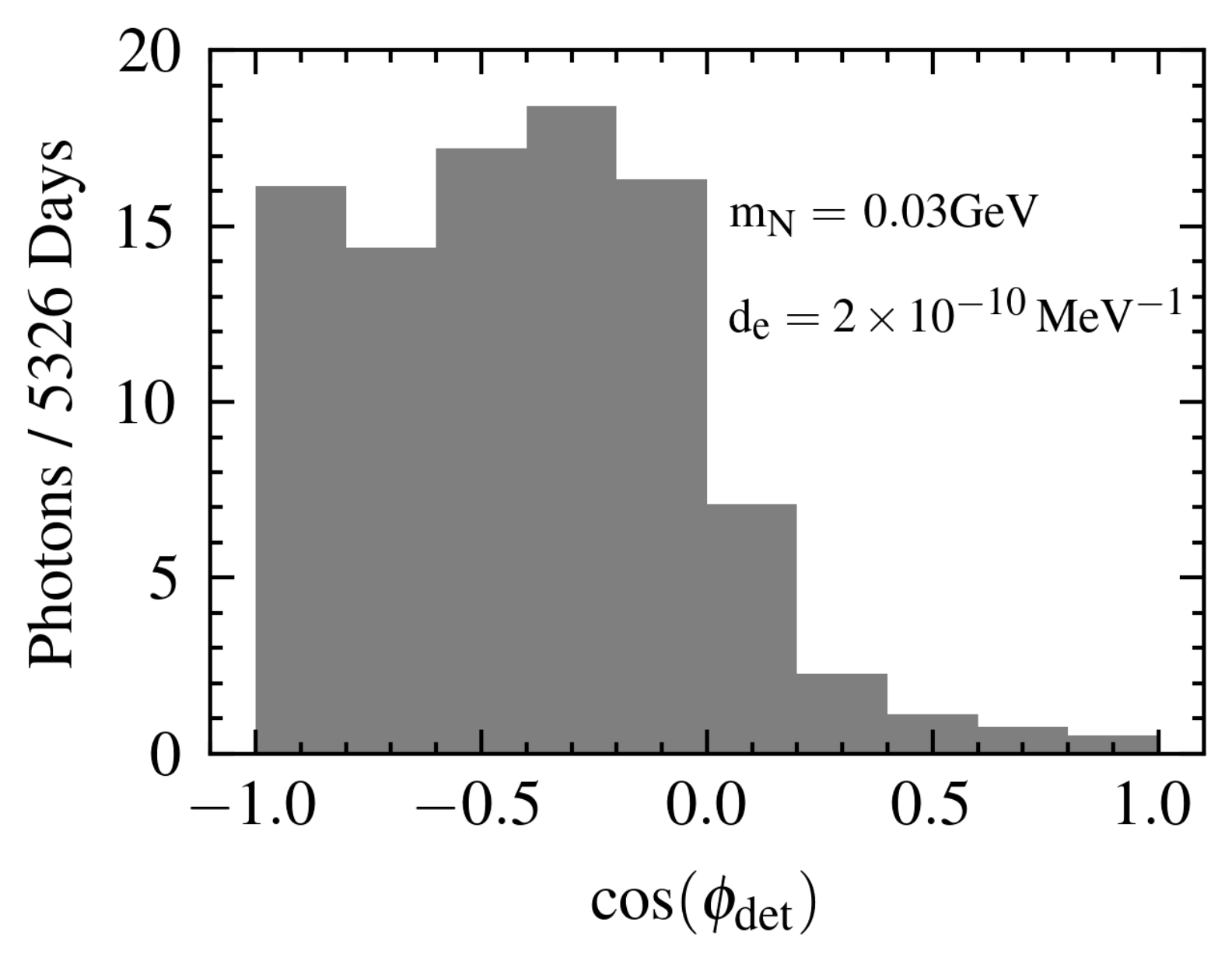}
    \caption{Angular distribution of the detected photons for Point A in \cref{Super_K_Plot}  ($m_N = 0.03$ GeV and $d = 2\times 10^{-10}\, \mathrm{MeV}^{-1}$). At low HNL masses and small dipole couplings, the angular distribution of photons is primarily up-going ($\cos\phi_{\mathrm{det}}<0$). This can be used in future searches to cut backgrounds. \label{Super_K_angles_lower}}
\end{figure}
We see that the photons are now more uniformly distributed in angle. This is because the decay length is now much shorter than the depth of the detector, so the volume available for upscattering is approximately spherical, and we expect our flux of HNLs at the detector to be roughly uniform in angle. We see that there is a slight peak near $\cos\phi_{\rm det} = 0$, since more neutrinos come from the horizontal direction than from the vertical.

The energy distribution for HNLs with shorter decay lengths (corresponding to large masses and strong couplings) is nearly uniform in the sub-GeV sample at Super-Kamiokande. This can be understood as follows: the flux of HNLs roughly mimics the flux of atmospheric neutrinos and so in the sub-GeV regime is relatively flat. The flux of photons is given roughly by $\dd N/\dd E_\gamma \sim  A_\perp \lambda P_{\rm vis}(\lambda, \ell)\int \dd E_N \Phi(E_N) {\rm Box}(E_\gamma| E_N)$ with $P_{\rm vis}(\lambda, \ell) =1-\exp(-\ell/\lambda)$. For long decay lengths, we find $\lambda P_{\rm vis} \approx \ell$ and the energy distribution is given by $\dd N/\dd E_\gamma \sim V_{\rm det} \times \int \dd E_N \Phi(E_N) {\rm Box}(E_\gamma| E_N)$. The box distribution of photons is flat, with a height that scales as $\sim 1/E_N$ such that the difference between the height of two bins of the histogram is given by
\begin{equation}\label{box-scaling}
    \Delta N_\gamma \approx \lambda P_{\rm vis}(\lambda, \ell) \times \Delta E_\gamma \times \frac{\Phi_N(E_N=E_\gamma)}{E_\gamma} ~.
\end{equation}
For $\lambda \gg \ell$, as in \cref{Super_K_energies_lower}, $P_{\rm vis} \approx \ell/\lambda$ and the overall rate is independent of $\lambda$ as discussed above. For $\lambda \sim \ell$, however, the probability of decaying inside the detector becomes some $O(1)$ number $P_{\rm vis} \approx 1- \exp[- \ell/\lambda]$ and the photon spectrum becomes proportional to $\lambda$. Since $\lambda\propto E_N \propto E_\gamma$, this cancels against the $1/E_\gamma$ denominator of \cref{box-scaling}, and the spectrum for shorter decay lengths inherits the shape of $\Phi_N$ which is relatively flat for sub-GeV energies; this explains the shape of \cref{Super_K_energies_upper}.

While our code is able to estimate the distribution of photons within Super-K, we only provide results from a rate-only analysis. A full analysis including the angular and spectral distributions of the decay photons would require considerations such as energetic and angular reconstruction of photon events in the detector, which are beyond the scope of this paper. We encourage those who have strong familiarity with Super-K to perform a more detailed analysis to improve constraints on new physics couplings.


We now consider future prospects and specifically upcoming experiments with larger fiducial volumes and stronger background rejection methods. We consider Hyper-K, DUNE, and JUNO, all with 10 years of live time. We assume that the $\frac{\rm Atm \, Rate}{\rm Fiducial \, volume}$ for all three of these experiments is the same as Super-K. We also assume that the energy range of interest will be the same as Super-Kamiokande ($30 {\rm ~MeV} ~- ~1.33 {\rm~GeV}$). When possible, we will consider an angular cut, meaning that we will only look at upward going events. For Super-K, the Monte Carlo predicts 51\% of the e-like 0-decay-e events to be down-going, while 49\% are up-going \cite{Super-Kamiokande:2019gzr}. We assume this holds for all experiments.

For ten years at Hyper-K, we expect roughly 70,000 sub-GeV $e$-like 0 decay-e atmospheric events. Super-Kamiokande run-IV is already doped with gadolinium and data from this exposure will have lower backgrounds from atmospheric neutrinos due to high-efficiency neutron tagging~\cite{Beacom:2003nk,Super-Kamiokande:2021jaq}. In our projections, we assume that Hyper-Kamiokande will be doped with gadolinium, which will cut the background from atmospheric neutrinos roughly in half. An angular cut will let us cut another $\sim 50\%$ of the background, leaving us with roughly 17,000 background events. Assuming no systematic uncertainty, a 95\%-CL bound can be set with 218 HNL events. This is, however, not likely to be realistic. The background uncertainties at Hyper-Kamiokande suffer from the same issues as at Super-Kamiokande where statistics are already high enough that a 5\% systematic uncertainty makes the search entirely systematics limited. The increased statistical sample will not be helpful unless the systematic uncertainty on the Monte Carlo prediction can be brought down to sub-percent levels even after accounting for reduced background rates from neutron tagging.


We expect roughly 9000 atmospheric events in a ``sub-GeV sample'' at DUNE over a ten year run-time. Since DUNE is a liquid-argon time projection chamber, it will be easier to distinguish the HNL decay products from a neutrino interaction. In particular, LArTPC technology offers: i) the ability to statistically discriminate between electrons and photons using measurements of $\dd E/\dd x$ at the beginning of a track, ii) MeV-scale reconstruction capabilities that can tag gamma rays from nuclear de-excitations \cite{ArgoNeuT:2018tvi}, and iii) the ability to measure final state charged hadrons including protons, pions with kinetic energies above $\sim 10$ MeV \cite{DUNE:2020lwj}. Finally, recent work has demonstrated that neutron tagging, using ``sprays'', may also be possible \cite{Friedland:2018vry}. Importantly, for our background tagging purposes we only need to veto nuclear scattering events and/or single electron showers which is a much easier task than the energy reconstruction considered in \cite{Friedland:2018vry}. While the ultimate capabilities of DUNE will require detailed simulation, we estimate that requiring upward going events will cut $50\%$ of the background, that proton tagging will catch $80\%$ of the remaining events, and that searches for neutron sprays and associated gamma rays from nuclear de-excitation can cut out $80\%$ of the remaining events for which no final state proton is produced. A naive combination of these estimates then suggests that 98\% of the background could be rejected at DUNE, however as we have already mentioned above, the precise value will require dedicated simulations. Performing an angular cut as well (requiring an up-going shower), we would expect a background of 88 events. With zero systematic uncertainty, 18 HNL event would set a 95\%-CL bound, and this conclusion remains unchanged even if we allow for a $\sim 5\%$ systematic on the background uncertainty. Note that unlike SK, DUNE will be statistically limited provided it can achieve background rejections that are better than $\sim 90\%$.

 \begin{figure*}
    \subfloat[Constraints assuming $d_e=d_\tau=0$ \label{subfig-mu-proj}]{%
      \includegraphics[width=0.475\linewidth]{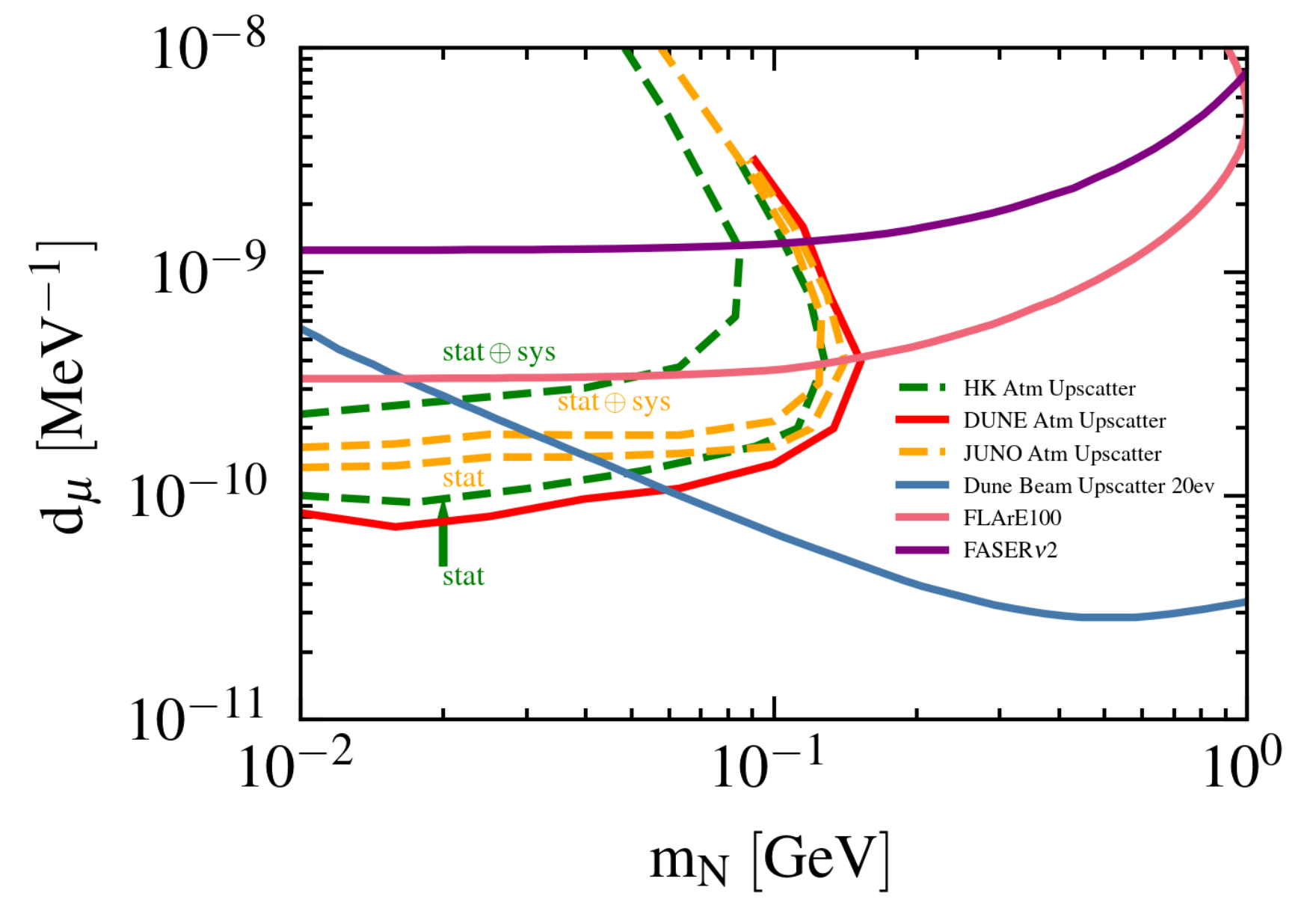}
    }
    \hfill
    \subfloat[Constraints assuming $d_\mu=d_\tau=0$ \label{subfig-e-proj}]{%
      \includegraphics[width=0.475\linewidth]{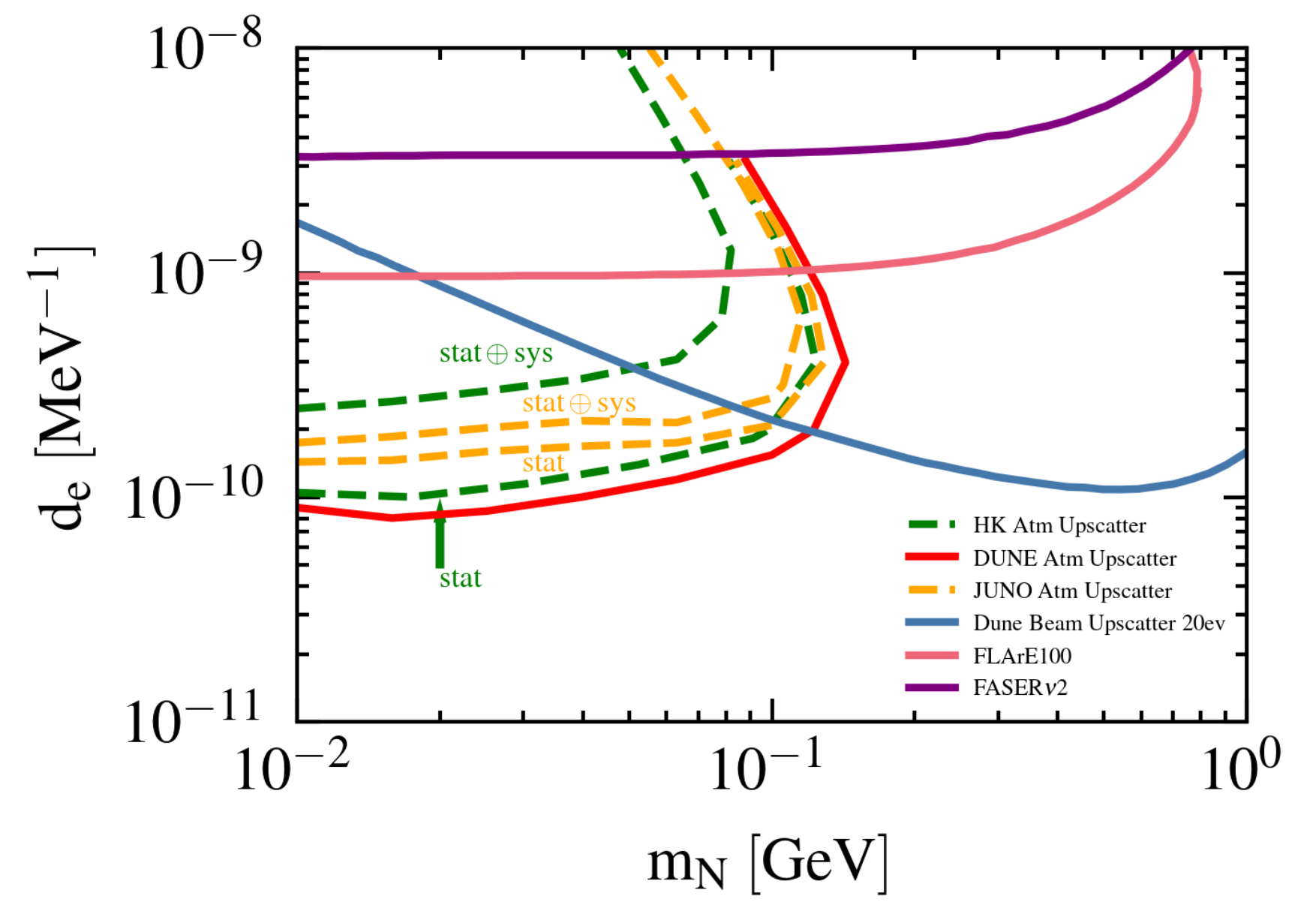}
    }
    
    \subfloat[Constraints assuming $d_e=d_\mu=0$ \label{subfig-tau-proj}]{%
      \includegraphics[width=0.475\linewidth]{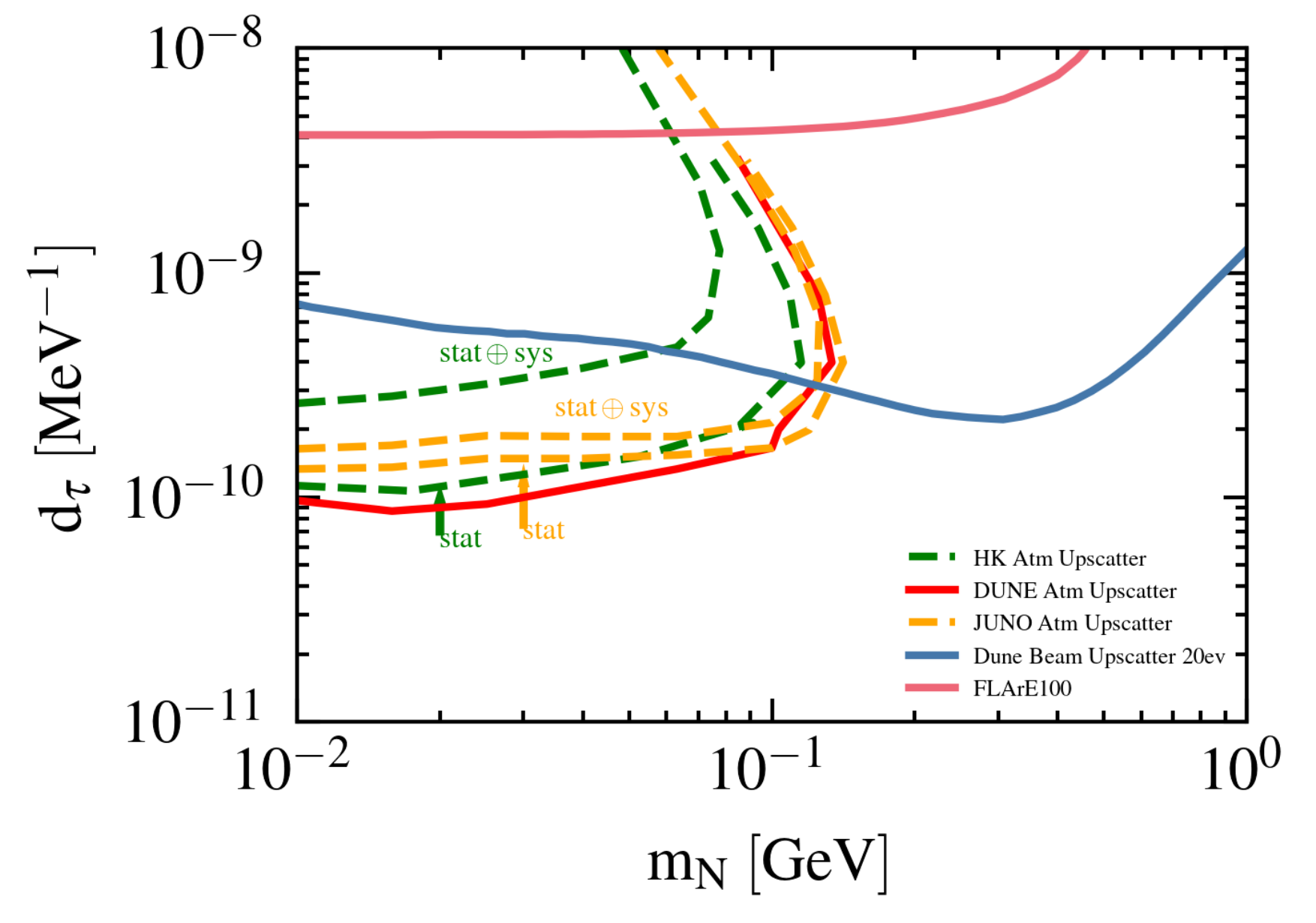}
      }
      \hfill
    \subfloat[Constraints assuming $U_{e,N}=U_{\mu,N}=0$ \label{subfig-mass-mix-proj}]{%
      \includegraphics[width=0.475\linewidth]{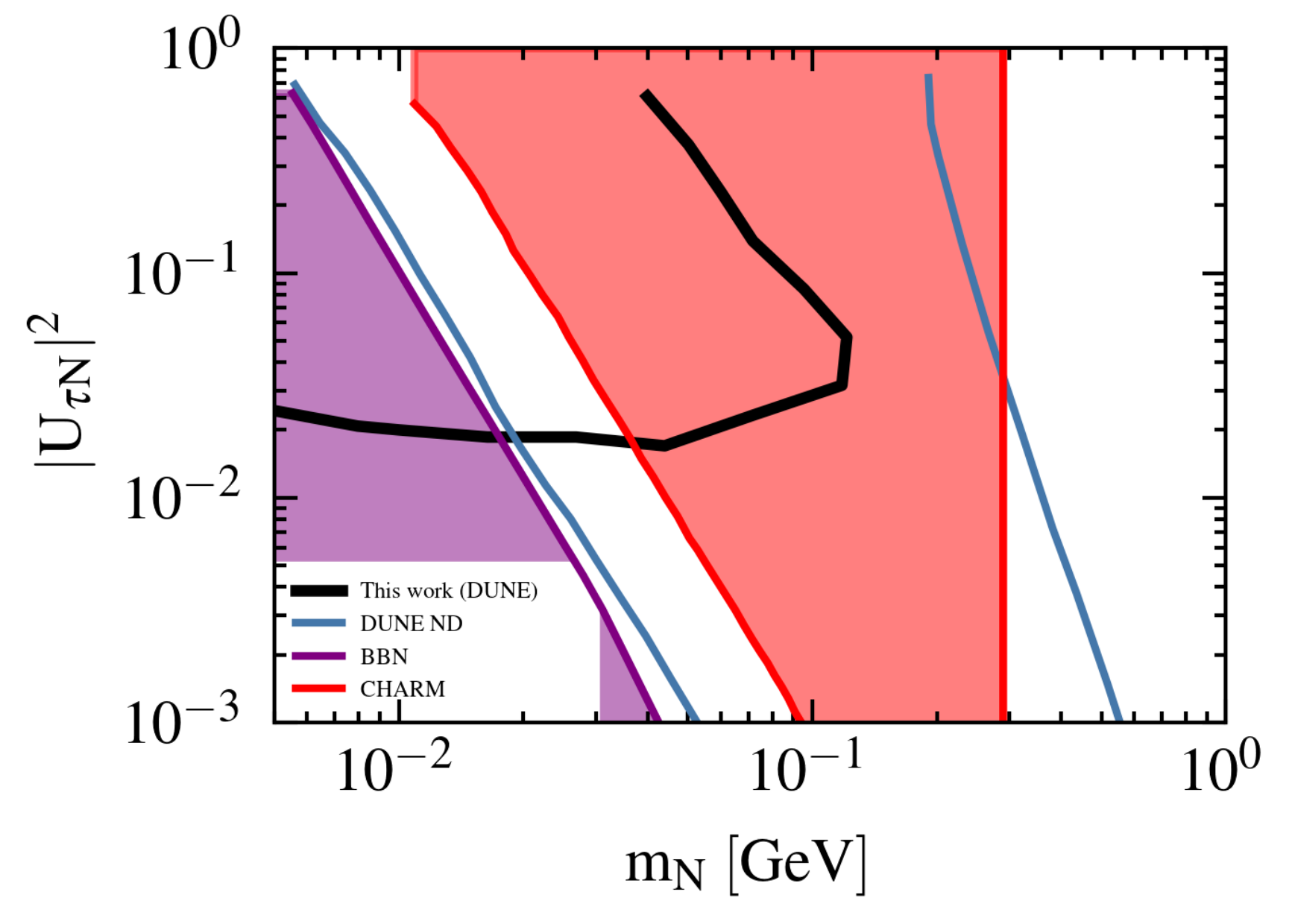}
      }  
    \caption{Comparison of our projections for future dipole-portal and mass-mixing-portal limits vs. other projections derived in the literature. Constraints from  FLArE 100 and FASER$\nu$2 were derived in \cite{Ismail:2021dyp}, and solar DUNE beam upscattering events were derived in \cite{Schwetz:2020xra}. We have included on Hyper-Kamiokande and JUNO for when $\sigma_{\rm{sys}} = 0$ and when $\sigma_{\rm{sys}}$ is 5\% of the projected background. For our DUNE projection, the background was so low that the two bounds were nearly identical, so we only included the $\sigma_{\rm{sys}} = 0$ curve. For the mass-mixing constraints, we use current bounds from \cite{Plestid:2020ssy,PhysRevLett.127.121801,Orloff:2002de,Boiarska:2021yho,Atkinson:2021rnp,sabti2020extended}, along with a projected bound for a multi-purpose DUNE near detector from \cite{Berryman:2019dme}.}
\end{figure*}

Finally, for JUNO, we estimate roughly 6,500  atmospheric events over 10 years. We expect that JUNO will be able to effectively cut out background from events that eject a proton due its low $\sim 1$ MeV detection threshold, and will have a 50\% efficiency at cutting background events that eject a neutron by leveraging the 2.2 MeV gamma ray from $ n p \rightarrow d \gamma$ \cite{JUNO:2015zny}. Using the relative distributions of neutrinos and anti-neutrinos, we estimate that 20\% of the atmospheric events will not produce a free proton and will instead produce a free neutron \cite{2019}. Assuming a $\gtrsim 95\%$ efficiency at tagging protons, we therefore estimate that the background from atmospheric neutrino CCQE will be $\sim 10\%$ of our re-scaled background from SK i.e.\ $\sim 650$ events. At these levels of statistics JUNO's statistical and assumed systematic uncertainties, taken again as $5\%$, are comparable.

For our projected limits on $|U_{\tau N}|^2$, we consider a ten year period at DUNE. As mentioned before, we expect to have significant background reduction by rejecting events that have evidence of scattering off of a nucleon. As a benchmark, we take 30 BSM events in ten years at DUNE to be statistically significant, and we show this contour in \cref{subfig-mass-mix-proj}. We see that the projected constraint from a multi-purpose near-detector at DUNE \cite{Berryman:2019dme} is quite close to constraints stemming from BBN (see \cref{subfig-mass-mix-proj}). These two constraints contain much of the parameter space that would be probed by our projected atmospheric upscattering method, and so while we expect improved sensitivity it is likely that the DUNE near detector can cover most of the relevant parameter space. It would be interesting for future studies to identify near-term experiments that may be able to supply new constraints using existing data sets by leveraging atmospheric upscattering.

\section{Conclusions \label{conclusions}} 

Atmospheric neutrinos have already given us a wealth of information regarding the nature of neutrino oscillation physics. Here we have seen that atmospheric neutrinos also provide a powerful and qualitatively distinct window into the nature of neutrino interactions and heavy sterile neutrinos. In particular, we have examined the upscattering of atmospheric neutrinos to HNLs and their subsequent decay inside of terrestrial detectors, finding that current data from Super-Kamiokande already yields leading constraints on both the dipole and mass-mixing portals. For the dipole portal, these bounds eat into new parameter space for HNL masses around $10~{\rm MeV} \lesssim m_{N} \lesssim 100~\rm{MeV}$, with precise constraints depending on which active neutrino flavor coupling dominates. Similarly, for the mass-mixing portal our Super-Kamiokande constraints provide leading constraints on the tau-sterile mixing angle for HNLs around $\sim 20~\rm{MeV}$. 

In the near future, experiments such as DUNE, Hyper-Kamiokande, and JUNO, will be able to take advantage of improved particle identification, background rejection, and employ dedicated search strategies incorporating angular distribution of the events to improve the bounds on the dipole couplings by around a factor of $\sim 2.5$ at low HNL masses. As such, this search strategy nicely complements the DUNE beam upscattering~\cite{Schwetz:2020xra} and double bang searches~\cite{Atkinson:2021rnp} which provide better sensitivity to higher HNLs masses, $m_{N} \gtrsim 100~{\rm MeV}$ on the dipole portal coupling, and HNLs produced from meson decays at DUNE for the mass-mixing portal~\cite{Coloma:2020lgy}. As an illustration of the strength of the future bounds, it is striking to observe that JUNO, Hyper-K, and DUNE appear poised to overlap in coupling reach with SN1987A~\cite{Magill:2018jla}, and will therefore close off an allowed gap in couplings.

\acknowledgements
     We thank Volodymyr Takhistov for comments on our manuscript and for useful discussions regarding the classification of single photons in SK. RP thanks Patrick Fox for comments on our manuscript and for helpful discussions regarding the numerical implementation of the upscattering volume integral, and Pedro Machado for comments on our manuscript and for discussion on atmospheric neutrino backgrounds in Super-Kamiokande and DUNE. We also thank Carlos Arg\"uelles for helpful discussions and feedback. This work was supported by the U.S. Department of Energy, Office of Science, Office of High Energy Physics, under Award Number DE-SC0019095 and  DE-SC002025. This manuscript has been authored by Fermi Research Alliance, LLC under Contract No. DE-AC02-07CH11359 with the U.S. Department of Energy, Office of Science, Office of High Energy Physics. This research was performed at the Kavli Institute for Theoretical Physics which is supported in part by the National Science Foundation under Grant No. NSF PHY-1748958 and at the Aspen Center for Physics, which is supported by National Science Foundation grant PHY-1607611.

\appendix

\section{Monte Carlo Routine \label{Monte-Carlo}}
\subsection{Sampling Values}
The Monte Carlo routine begins by sampling a neutrino energy $E_{\nu}$. We want our distribution of energies to follow a power spectrum, since both the flux of atmospheric neutrinos and the scattering cross-section follow a power law with respect to energy. We consider a maximum (minimum) energy $E_{\nu, \rm max} (E_{\nu, \rm min})$,
\begin{equation}
    \rho_E = \kappa E_{\nu}^{-\gamma}~,
    \label{eq:Energy_Sample}
\end{equation}
\begin{equation}
    \kappa = \frac{1 - \gamma}{E_{\nu, \rm max}^{1-\gamma} -E_{\nu, \rm min}^{1-\gamma}}~.
\end{equation}
To get our energy, we uniformly sample a number $\chi \in [0,1]$ and then
\begin{equation}
    E_{\nu} = \bigg ( \frac{1-\gamma}{\kappa}\chi + (E_{\nu, \rm min})^{1-\gamma} \bigg )^{\dfrac{1}{1-\gamma}}
\end{equation}

To sample the position of interaction, we want a distribution that mimics the decay length of the HNL. We define our minimum desired distance from the detector as $R_{\mathrm{min}}$, and the maximum distance $R_{\mathrm{max}} = R_{\oplus} + |\mathbf{Y}|$. Then, 
\begin{equation}
    \begin{split}
        r' =& R_{\mathrm{min}} - \lambda \ln\bigg( 1 - \chi \qty[ 1 - \e^{\frac{\Delta R}{\lambda} }] \bigg) ~,
    \end{split}
\end{equation}
with $\Delta R=R_{\rm min} - R_{\rm max}$, and $\chi \in [0,1]$. We then sample $\phi'$ uniformly, and sample $\cos\theta'$ uniformly from angles that leave the interaction within the Earth. Then, the interaction location is
\begin{equation}
    \vb{X} = \vb{Y} + \bigg ( r' \sin \theta'\cos\phi', r' \sin\theta'\sin\phi' , r' \cos\theta' \bigg )
\end{equation}
If $\vb{X} > R_{\oplus}$, we reject and repeat the sampling until we get an interaction position within the Earth.

For sampling the scattering angles, we note that some models favor forward scattering, while others have scattering cross-sections that have quasi-uniform angular distributions. We define a minimum (maximum) scattering angle $\Theta_{\rm min}$ ($\Theta_{\rm max}$) based on allowed kinematics, and a value of $\epsilon \in \{0,1\}$ based on the type of scattering
\begin{equation}
    \rho_{\Theta} = \beta (1 - \cos\Theta)^{-\epsilon}~,
\end{equation}
so that for $\epsilon=1$ we have
\begin{align}
    \beta &= \big[ \ln \big( (1-\cos\Theta_{\max})/(1-\cos\Theta_{\min}) \big) \big] ^{-1}~,
\end{align}
while for $\epsilon=0$ we have 
\begin{align}
    \beta = [\cos\Theta_{\min} - \cos\Theta_{\max}]^{-1} ~.
\end{align}
For sampling the scattering angle, we sample a uniform number $\chi \in [0,1]$, such that the angles are sampled as
\begin{align}
    \cos\Theta &= 1 - (1 - \cos\Theta_{ \rm max} )^{\chi} (1 - \cos\Theta_{\rm min})^{1-\chi} \\
    \cos\Theta &= \cos\Theta_{\max} + \chi (\cos\Theta_{\min} - \cos\Theta_{\max}) ~,
    \label{eq:Cos_Theta_Sample}
\end{align}
for $\epsilon=1$ and $\epsilon=0$ respectively. 

Finally, we want to sample the neutrino entry position $\vb{W}$. We define $\vb{v}_{\rm in} = \vb{X} - \vb{W}$. Using our scattering angle $\Theta$ that we sampled and another angle $\psi$ uniformly sampled from 0 to 2$\pi$, then

\begin{equation}
    \begin{split}
    \hat{\vb{v}}_{\rm in} = \frac{\vb{Y} - \vb{X}}{|\vb{Y} - \vb{X}|} \cos\Theta +& \vb{\hat{v}_{1\perp}} \sin\Theta \cos\psi \\
    +& \vb{\hat{v}_{2\perp}} \sin\Theta \sin \psi
    \end{split}
\end{equation}
Where $\vb{Y} - \vb{X}$, $\vb{v_{1 \perp}}$, $\vb{v_{2\perp}}$ are all mutually orthogonal. To find the length of the path travelled, we use
\begin{equation}
    |\vb{v_{in}}| = \vb{X} \cdot \vb{\hat{v}_{in}}+ \sqrt{(\vb{X} \cdot \vb{\hat{v}_{in}})^2 + R_{\oplus}^2 - |\vb{X}|^2}
\end{equation}

Finally, we get $\vb{W} = \vb{X} - \vb{\hat{v}_{in}} |\vb{v_{in}}|$

\subsection{Calculations}
    Having sampled our energy, scattering angle, interaction position, and neutrino entry position, we can calculate other necessary values. From $\vb{W}$ and $\vb{X}$, we calculate the zenith angle $\phi_{\rm zen}$ of the incoming neutrinos (\cref{cartoon}). We use \texttt{NuFlux}  \cite{NuFlux:2022} to calculate $\mathcal{I}_{\rm incoming}(E_{\nu},\phi_{\rm zen})$. When working with flavor dependent couplings, we calculate a 1D density profile from $\vb{W}$ to $\vb{X}$ using the Preliminary Earth Reference Model (PREM) \cite{Dziewonski:1981xy}. Oscillations are calculated by integrating along this density profile to obtain $I_{\nu_{\alpha}}(E_{\nu}, \phi_{\rm zen}, \vb{X})$.

At $\vb{X}$, we use the PREM \cite{Dziewonski:1981xy} to calculate the density, and call a saved dictionary of the number density for each of the elements \cite{MCDONOUGH2014559, lutgens2011essentials}. We calculate the cross sections for the scattering using methods described in \cref{Cross_Sections} to get $\sum_A n_A \frac{\dd\sigma_A}{\dd \cos \Theta} = (n \frac{\dd \sigma}{\dd \cos \Theta})_{\rm eff}$. We  calculate the decay length of the HNL using \cref{eq:Gen_Decay_Length}, and probability for producing a visible decay from \cref{eq:P_Dec}

\subsection{Weighting}
Since we preferentially sample our values, we must include a weighting when calculating the rate of decays from the Monte Carlo. These weights are calculated by taking the ratio of our sampling distribution to the true integrand. Explicitly, the weights are given by,
\begin{align}
    w_E(E_{\nu, i}) &= \frac{(E_{\nu,i})^{\gamma}}{\kappa (E_{\nu,{\rm max}} - E_{\nu,{\rm min}})}\nonumber ~,\\
    w_{\Theta}(\Theta_i) &= \frac{1}{\beta} \frac{(1 - \cos\Theta_i)^{\epsilon}}{\cos(\Theta_{\rm min}) - \cos(\Theta_{\rm max})}  ~,     \label{Sampling_Weights} \\
    w_V(\lambda_i,r'_{i}) &= \frac{\e^{r'_i/\lambda}\qty[\e^{- R_{\mathrm{min}}/\lambda_i} -\e^{- R_{\mathrm{max}}/\lambda_i}]}{\big( R_{\mathrm{max}} - R_{\mathrm{min}}\big) V_{\oplus}} \frac{\mathrm{d}V}{\mathrm{d}R} R_{\max}\nonumber~,
\end{align}
where
\begin{equation}
    \frac{\mathrm{d} V}{\mathrm{d}R} = \begin{cases} 
    &\frac{4 \pi}{3} (r')^2 \, \, \mbox{if} \, \, R_{\oplus} > |\mathbf{Y}| + r' \\
    &\frac{\pi r'}{|\mathbf{Y}|} \big( R_{\oplus}^2 - [|\mathbf{Y}| - r']^2 \big)~ \mbox{else}~.
    \end{cases}
\end{equation}
In a Monte Carlo without preferential sampling, we would have standard ``Reimann weights'', $\Delta\textbf{X} \Delta E_{\nu} \Delta\cos\Theta/N$ with $\Delta \textbf{X} = V_{\oplus}$,  $\Delta E_{\nu}=(E_{\nu, \rm max}-E_{\nu, \rm min})$,  $\Delta\cos\Theta= \big( \cos(\Theta_{\min}) - \cos(\Theta_{\max}) \big)$, and $N$ the number of samples in our Monte Carlo. Notice that upon combination (multiplying all weights together) the denominators in our new sampling weights, \cref{Sampling_Weights},  cancel against the uniform weights such that  we obtain the weight for the $i^{\rm th}$ sample as
\begin{equation}
    \begin{split}
        w_i=
        \frac{V_{\oplus}(E_{\nu,i})^{\gamma} (1 - \cos\Theta_i)^{\epsilon}}{\kappa \beta N}  w_V(\lambda_i, r'_i)~.
    \end{split}
    \label{eq:Pref_Deltas}
\end{equation}

We now have everything needed to compute the rate
\begin{equation}
    \begin{split}
        R = &\sum_{i = 1}^{N} \bigg(n \frac{\dd \sigma}{\dd \cos\Theta}\bigg)_{\rm eff} \times \mathcal{I}_{\nu}\times \frac{P_{\rm vis}}{4 \pi |\vb{X} - \vb{Y}|^2} w_i  ~.
    \end{split}
    \label{eq:Master_Sum}
\end{equation}
This is the numerical cousin of \cref{master-eq}.  In this routine, between 20,000 and 300,000 events were generated for each simulation.  The larger simulations were necessary for the mass-mixing model.

\section{Upscattering Cross sections \label{Cross_Sections}}
\subsection{Dipole Portal}
For dipole upscattering, the cross section can be decomposed into
\begin{equation}
    \begin{split}
        \frac{\dd \sigma}{\dd t} = \frac{\dd \sigma_{\rm coh}}{\dd t}& |F(-t)|^2 \\
        +& \frac{\dd \sigma_{\rm p}}{\dd t} Z (1 - |F(-t)|^2 ) + \frac{\dd \sigma_{\rm n}}{\dd t} (A - Z) ~,
    \end{split}
    \label{eq:Full_Cross_Section}
\end{equation}
where $A$ and $Z$ are the atomic mass and atomic number of the nucleus respectively and $F(-t)$ is the form-factor for the transferred momentum. We work in the infinite mass limit, so our transferred momentum goes as
\begin{equation}
    t_{\rm coh} = 2 E_{\nu}^2 - m_N^2 - 2 E_{\nu} \sqrt{E_{\nu}^2 - m_N^2} \cos\Theta ~.
    \label{eq:T_From_Theta_Coherent}
\end{equation}
The coherent scattering is given in \cref{eq:Coh_Cross_Sec} with $\vb{Q}^2 = -t$. Meanwhile, incoherent scattering off of protons or neutrons will go as
\begin{align}
        &\frac{\dd \sigma_{\mathrm{in}}}{\dd t} = \frac{\alpha d^2 (m_N^2 t - 2 m_N^4 + t^2)}{m_p^2 t^2 (m_p^2 - s)^2} \\
        &~~~~\times \qty[2 F_1^2 m_p^2 (2 m_p^2 + t) - 12 F_1 F_2 m_p^2 t + F_2^2 t (8m_p^2 + t) ]~.\nonumber
\end{align}
Here, $m_p$ is the mass of a nucleon, and s is the center-of-mass energy given by $m_p^2 +2 m_p E_{\nu}$. The value of t for incoherent scattering goes as
\begin{equation}
    t_{\mathrm{in}} = m_N^2 - 2 E_{\nu} \big( E_N - \sqrt{E_N^2 - m_N^2} \cos\Theta \big)~.
    \label{eq:T_From_Theta_Incoherent}
\end{equation}
If the neutrino scatters incoherently, then the energy of the HNL is

\begin{align}
    &E_{N \, \rm incoh} = \frac{(E_{\nu}+m_p) (m_N^2 + 2 E_{\nu} m_p)}{2 \big( E_{\nu}^2 \sin^2\Theta + 2 E_{\nu} m_p + m_p^2 \big)}\\
    &+\frac{E_{\nu}\cos\Theta \sqrt{-4 m_N^2 \big( m_p^2 + E_{\nu}^2 \sin^2 \Theta \big) + m_N^2 - 2 E_{\nu} m_p^2}}{2 \big( E_{\nu}^2 \sin^2\Theta + 2 E_{\nu} m_p + m_p^2\big )}\nonumber
    \label{eq:E_N_incoherent}
\end{align}

Finally, we need to calculate the $F_1$ and $F_2$ values \cite{Borah:2020gte} \cite{PhysRevC.69.022201}

\begin{equation}
    \begin{split}
        F_{1,p/n} &= \frac{1}{1 - \frac{Q^2}{4 m_p^2}} \qty(G_E^{p/n} + \frac{Q^2}{4 m_p^2} \times G_M^{p/n}) \\
        F_{2,p/n} &= \frac{1}{1-\frac{Q^2}{4 m_p^2}} \qty(G_M^{p/n} - G_E^{p/n}) ~,
    \end{split}
\end{equation}

where
\begin{equation}
    \begin{split}
        G_E^p &= G_D\\
        G_E^n &= 0\\
        G_M^{p,n} &= \mu_{p,n} G_D\\
        G_D &= \bigg( 1 + \frac{Q^2}{0.71 \mathrm{GeV}^2} \bigg)^{-2},
    \end{split}
\end{equation}
and $\mu_p$ = 2.793, $\mu_n$ = -1.913 and $Q^2 = t$.

In our routine, while we use \cref{eq:T_From_Theta_Coherent} and \cref{eq:T_From_Theta_Incoherent} for finding the transferred momentum, we do not know if the true scattering is coherent or incoherent (\cref{eq:Full_Cross_Section} says that we have components of both). Therefore, we let the energy of the propagating HNL be $E_{\nu}$.

When implementing the full cross section, we find that the coherent part still dominates, so most of our phenomenology can be explained by considering the coherent case.

\subsection{Mass Mixing Routine}
In the mass-mixing portal, we must consider coherent elastic scattering, incoherent scattering on nucleons, and deep-inelastic scattering (DIS). Unlike the dipole portal, where nearly all scattering is coherent, the mass-mixing model has significant contributions from incoherent scattering. Since different forms of scattering leads to different HNL energies (and therefore different observed energies), we run the simulations independently for each type of scattering, and then sum together the rates to get the total contribution.

For coherent scattering, we have

\begin{equation}
    \frac{\dd \sigma_{\mathrm{coh}}}{\dd t} = \frac{|U_{\alpha N}|^2 G_F^2 Q_w^2}{2 \pi} \bigg( 1 - \frac{m_N^2}{4 E_{\nu}^2} + \frac{t}{4 E_{\nu}^2} \bigg) |F(-t)|^2
    \label{eq:MM_coh_sigma}
\end{equation}

In \cref{eq:MM_coh_sigma}, $G_F$ is the Fermi coupling constant and $Q_w$ is the weak charge of the nucleus. 

Incoherent scattering is modelled by treating nuclei as collections of free nucleons and using standard hadronic form factors for the nuclei. We take dipole parameterizations of the vector, magnetic, and axial form factors and rely on a partially conserved axial current ansatz for the pseudoscalar form factor; explicit expressions can be found in \cite{Formaggio:2012cpf,LlewellynSmith:1971uhs}.

For DIS, we consider scattering off of individual quarks. To find the cross section, we first find cross section for scattering off of quarks $\sigma_f$ as a function of the momentum carried by the quark. This is parameterized by $x$, the fraction of the total longitudinal nucleon momentum carried by the quark. Finally, we have,

\begin{equation}
    \sigma_{\mathrm{DIS}} =\int_0^1 \bigg(\sum_f \sigma_f(x) f_f(x) \bigg) \dd x,
    \label{eq:Calc_DIS}
\end{equation}
where $f_f(x)$ is the parton distribution function (PDF) for the particular quark.

We numerically performed the integral in \cref{eq:Calc_DIS}, using PDFs from \cite{Hou:2019efy} and treating the HNL as massless, since the masses we are sensitive to are far below the GeV energy scale where DIS becomes important. Although the cross section for DIS is scales linearly with neutrino energy the DIS scattering only contributes on the order of a few percent in the region of parameter space that is not covered by existing searches. Therefore, in our code, we use a simple form for DIS with the leading coefficient determined by \cref{eq:Calc_DIS},

\begin{equation}
    \begin{split}
    \frac{\dd \sigma_{\mathrm{DIS}}}{\dd \cos\Theta} = &\frac{|U_{\alpha N}|^2 E_{\nu} }{2}  \sqrt{1 - \frac{m_N^2}{s}} \\
    &\times(1 - |F(-t)|^2)\times 3 \times 10^{-39} \mathrm{cm}^2~.
    \label{eq:MM_DIS_sigma}
    \end{split}
\end{equation}

From these scattering channels, we can determine the number of visible decays (in this case, $N \rightarrow \nu e^{+} e^{-}$) expected for Super-K. We assume that Super-Kamiokande will be unable to resolve both the $e^{+}$ and the $e^{-}$, so the decay will appear as a sub-GeV 0 decay-e event. We calculate the energy of the $e^{+} e^{-}$ pair by using the invariant mass distribution in \cite{de2021three}, and require that the energy be between 30 MeV and 1.33 GeV. We see the resulting bounds in \cref{subfig-U}.

\subsection{Form Factor Fitting \label{form-factors}}

We can see that for \cref{eq:Full_Cross_Section}, we need a way to calculate the nuclear form factor. The Helm form factor allows us to accomplish this
\begin{equation}
    F_{\rm Helm}(Q^2) = \bigg( \frac{3 j_1(Q R_1)}{Q R_1} \bigg)^{1/2} \times\e^{ - (s Q)^2/2}~,
    \label{eq:Helm_Form_Factor}
\end{equation}
where $j_1$ is a spherical Bessel function of the first kind and
\begin{equation}
    R_1 = \sqrt{R_A^2 + \frac{7 \pi^2}{3} r_0^2 - 5 s^2}~.
\end{equation}

Rather than using default parameters as a global description of all nuclei, we fit the values of $R_A$, $r_0$, and $s$ independently for each nucleus, and then store these values for later calculation. We begin by taking the 2-parameter Fermi distribution for the radial nuclear charge distribution from \cite{DeVries:1987atn}. Taking the 3D Fourier transform of this charge distribution gives us the charge form factor. Using initial values of $R_A$, $r_0$, and s, we define the difference between the Fermi and Helm form factors as
\begin{equation}
    S = \sum_i \big( F^2_{\rm Helm}(Q_i^2, R_A, r_0, s) - F^2_{\rm Fermi}(Q_i^2) \big)^2~.
\end{equation}
We then use gradient ascent to iteratively improve our fit (i.e. $R_1 \rightarrow R_1 - (\dd S/ \dd R_1) \delta R_1$ where $\delta R_1$ is some predetermined constant).

\section{Earth Model   \label{Earth Model}}

To find the local density of the the Earth, $\rho(X)$, at the location of neutrino interaction, we use the Preliminary Reference Earth Model, specifically Table IV of \cite{Dziewonski:1981xy}.

We need to determine the number density of each element at the interaction location. To do this, we obtain the elemental weight fraction of the core and mantle from \cite{MCDONOUGH2014559}, and the crust weight fractions from \cite{lutgens2011essentials}. The results are summarized in \cref{table:weights}

\begin{table}
\centering
$\begin{array}{ cccc}
\text{Element} 		& \text{Crust \%}  & \text{Mantle \%} & \text{Core \%} 	 \\
\midrule


\text{O} & 46.6 &  44 & 0 \\ 
\text{Si} & 27.72 &  21 & 6 \\ 
\text{Al} & 8.13 &  2.35  & 0 \\ 
\text{Fe}  & 5.05 & 6.26 &  85.5 \\
\text{Ca} & 3.65 &  2.5 & 0 \\ 
\text{Na} & 2.75 & 0 & 0 \\
\text{K} & 2.58 & 0 & 0 \\
\text{Mg} & 2.08 & 22.8 & 0 \\
\text{S} & 0 & 0 & 1.9 \\
\text{Ni} & 0 & 0 & 5.2 \\
\midrule
\midrule
\text{Total} & 98.56 & 98.91 & 98.6 \\
 \bottomrule
\end{array}$

\caption{Fractional weights for elements in each layer of the Earth
\label{table:weights}}
\end{table}

We can now calculate the number density of each element $n_i$ by

\begin{equation}
    n_i(\mathbf{X}) = \frac{\rho(\mathbf{X}) f_i}{m_i} \times  N_A,
\end{equation}

where $N_A$ is Avagadro's number, and $f_i$ and $m_i$ are the fractional weight and molar mass of the element in question, respectively.
\bibliography{biblio.bib}

\end{document}